\documentclass[a4paper,12pt]{article}
\usepackage{jheppub} 
\usepackage{lineno}

\usepackage[english,activeacute]{babel}
\usepackage{mathtools}
\usepackage[scr=boondox,  
            cal=esstix]   
           {mathalpha}

\title{\boldmath Kerr-Schild solutions in Multigravity and the Classical Double Copy}

\author[1]{Hugo Garc\'{\i}a-Compe\'an}

\author[2]{and Everardo Rivera-Oliva}
\affiliation{Departamento de Física,\\
Centro de Investigaci\'on y de Estudios Avanzados del Instituto Polit\'ecnico Nacional,\\
P.O. box 14-740, C.P. 07000, Ciudad de México, Mexico}
\emailAdd{hugo.compean@cinvestav.mx}
\emailAdd{everardo.rivera@cinvestav.mx}

\abstract{We explore the multimetric theory of gravitation, also known as multigravity. We derive additional new exact solutions for the theory in proportional Kerr-Schild and double Kerr-Schild forms. We extend several solutions from the theory of General Relativity, characterized by a constant Ricci scalar in single and double Kerr-Schild forms, to derive solutions in the multi-gravity context. We also examine and extend the classical double copy relations that can be constructed out from these solutions in multigravity exploring the dynamics of the single copy and zero copy fields.}

\makeatletter
\gdef\@fpheader{}
\makeatother

\begin{document}
\maketitle
\flushbottom
\section{Introduction}
The path to quantum gravity has several subtleties and candidates (see \cite{Armas:2021yut} for a review). One of them modifies Einstein's theory of General Relativity (GR) at the classical level due to effective corrections coming from quantum effects. From the perspective of particle physics, we think of the gravitational field as a spin-2 field with massless excitations known as gravitons that mediate the gravitational interaction. In fact, it can be shown that that GR is the unique theory that can be constructed from a spin-2 massless field theory with local interactions, which is Lorentz invariant and unitary, as can be seen in \cite{Feynman:1996kb,Gupta:1954zz,Wald:1986bj,Weinberg:1965rz,Weinberg:1980kq}. 

It is also possible to examine this process in reverse, beginning with an altered classical gravitation theory and tackling the physical phenomena that GR fails to explain. The fundamental problems of dark matter and dark energy are some of the sources of the need for a modified theory of gravity. Modifications to GR include Lorentz violating and non-local theories (see \cite{Mattingly:2005re,Deser:2007jk}); however, the conservation of Lorentz symmetry has ever been found to be valid experimentally. Another possible modification consists of a theory with a massive graviton. Massive gravity has been explored since the work of Fierz and Pauli \cite{Fierz:1939ix}, where an additional mass term was introduced in the perturbation around the background metric, resulting in a linearized version of a theory of massive gravity. The non-linear extension of the Fierz-Pauli theory leads to two issues: the van Dam–Veltman-Zakharov (vDVZ) discontinuity \cite{vanDam:1970vg,Zakharov:1970cc} and the Boulware-Deser (BD) ghost problem \cite{Boulware:1972yco,VanNieuwenhuizen:1973fi}. 

For the Boulware-Deser problem, a theory of a spin-2 massive field with six degrees of freedom exists. One of these modes is a scalar mode known as the BD ghost. The energy of the BD ghost is negative, which leads to an unstable vacuum, making the theory inconsistent. A Boulware-Deser ghost free theory of massive gravity was formulated by de Rham-Gabadadze-Tolley and is commonly known as dRGT gravity (see \cite{deRham:2010kj}). In dRGT gravity there is a fixed spin-2 metric field that interacts with the GR metric field. The discontinuity of vDVZ is fixed in this theory by the Vainshtein mechanism \cite{Vainshtein:1972sx}. A massive spin-2 metric field propagates 5 degrees of freedom; however, through the Vainshtein mechanism, this extra degree of freedom responsible for the vDVZ discontinuity gets screened by its own interactions. 

The consideration of massive gravity has resulted in the development of bimetric gravity, a theory with two general metrics, one of them is dynamical, and the other is a reference metric. This extension for two gravitational fields (see \cite{Hassan:2011zd}) involves the interaction of two dynamic spin-2 fields (for some important reviews on massive gravity and bigravity, see \cite{Hinterbichler:2011tt,deRham:2014zqa,Schmidt-May:2015vnx}). This has further evolved into multimetric gravity theory, also referred to as multigravity, which describes the interaction among $N$ dynamic spin-2 fields. All these generalizations of massive gravities into interacting spin-2 fields can be obtained from the deconstruction paradigm (see \cite{deRham:2014zqa}). Multigravity was an alternative proposal to GR first motivated by trying to explain the origin of the late time accelerated expansion of the universe \cite{intro1,intro2}, but later was found useful to explore dark matter \cite{GonzalezAlbornoz:2017gbh,Babichev:2016bxi,Babichev:2016hir,Marzola:2017lbt} and the hierarchy problem \cite{Niedermann:2018lhx,Avgoustidis:2020wrd} by the introduction of additional massive spin-2 gravitons above the massless graviton of GR. In the search for solutions in multigravity, some black-hole-like solutions have been considered \cite{Hassan:2012wr, Baldacchino:2016jsz}. These solutions are of the proportional type, in which all metrics are proportional to the one of GR differing only in a constant conformal factor. Non-proportional (non-diagonal) solutions also arise in this context with a more complicated structure and properties. It is known that multigravity solutions present certain instabilities for some values of the graviton mass \cite{Babichev:2013una,Brito:2013wya}. Moreover, some of the non-proportional black holes will be stable \cite{Babichev:2014oua,Babichev:2015zub} (for a more detailed account on this topic, see \cite{Babichev:2015xha}). For simplicity in the present work we will focus only on solutions of the proportional type.

On the other hand, in Refs. \cite{Bern:2008qj,Bern:2010ue,Bern:2010yg} it was proved that the $n$-leg tree-level amplitude in Yang-Mills theory can be expressed as a summation over the amplitudes of cubic diagrams. Although Yang-Mills theory includes interaction diagrams with both cubic and quartic vertices, one can always restructure the color factors to describe quartic-vertex diagrams as a sum of cubic-vertex diagrams \cite{Bern:2008qj,Bern:2010ue,Bern:2010yg}. The color factors in the gauge amplitudes are configured such that they encompass the weights of the rearranged quartic-vertex diagrams. Thus, once the Yang-Mills amplitude is expressed solely as a sum of cubic diagrams, they fulfill the color-kinematics duality, or the BCJ duality \cite{Bern:2008qj,Bern:2010ue,Bern:2010yg}. Then the color and the kinematic factors have a very similar structure (Jacobi identity) and can be exchanged. Using BCJ duality it becomes feasible to derive a $n$-leg amplitude in gravitational theory by replacing the color factor in the gauge amplitudes with an alternative kinematic factor and substituting the coupling constant of the gauge theory with the gravitational coupling constant. This method is referred to as the double copy procedure. Although it is not necessary for the color factor to be replaced with a kinematic factor of the same theory, as demonstrated in \cite{Bern:2008qj,Bern:2010yg}, this paper will restrict its scope to such instances.  The correctness of this procedure has been demonstrated based on the KLT relations as formulated in \cite{Kawai:1985xq}, extending up to the eighth order as outlined in \cite{Bianchi:2008pu}, and it has been rigorously proven to be valid at all orders at the tree level in \cite{Bern:2010yg}. It is postulated that the procedure remains applicable at the loop level, a notion that Bornsten, White, and others have investigated in \cite{Bern:1998sv,Borsten:2020zgj,Borsten:2021gyl,White:2017mwc}.

The double copy map has been also considered for the case of the field equation's solutions both in gauge and gravity theories. This is termed the {\it classical double copy}. This is a map of solutions in both theories \cite{Monteiro:2014cda}. For a very complete and recent review on this topic, see \cite{White:2024pve}. The first consideration of this type of theories works very well for the Kerr-Schild ansatz \cite{Kerr:1965wfc}. Some of the applications of this version of the double copy were carried out in Refs.  \cite{Luna:2015paa,Luna:2016due,Ridgway:2015fdl,Bahjat-Abbas:2017htu,Carrillo-Gonzalez:2017iyj,CarrilloGonzalez:2019gof,Berman:2018hwd}. More recently, an alternative (and more general) proposal was introduced, the so-called Weyl double copy \cite{Luna:2018dpt}.  This last proposal has been shown to be strongly related to twistor theory  \cite{White:2020sfn,Chacon:2021wbr,Chacon:2021lox,Luna:2022dxo}. However, for simplicity, in this work, we will focus on the Kerr-Schild double copy.

In the present article, we explore the multi-metric theory of gravitation. In particular, we derive additional exact solutions in multigravity in Kerr--Schild and double Kerr--Schild forms directly in the metric formalism for the first time, in contrast to previously known multigravity solutions that were obtained using the vielbein formalism. We will focus on extending the classical double copy relations in the context of multigravity by deriving exact solutions in Kerr-Schild and double Kerr-Schild forms in the multimetric theory of gravitation and connecting them with solutions in multi field gauge theory. We build on work done in the context of the double copy description of bigravity carried out in \cite{Garcia-Compean:2024uie,Garcia-Compean:2024zze,Garcia-Compean:2025wkj}. Moreover, we explore this connection at the theoretical level. We investigate which multi-field gauge theory underlies the single-copy and zero-copy maps, namely a quadratic $U(1)^N$ Proca (multi-vector) theory for the single copy and an interacting multi-scalar theory (with $SO(3)^N$ symmetry) for the zero copy, both of which have direct applications in cosmology (dark-photon sectors \cite{Fabbrichesi2021} and multifield inflation \cite{PhysRevD.110.L041302,Wands2007}).

The work is organized as follows: In Section \ref{S2} we present a brief overview of the Boulware-Deser ghost free theories of massive gravity and its generalizations including bimetric gravity and multigravity. We also derive the equations of motion for multigravity. In Section \ref{S3}, we build solutions in multigravity by using proportional Kerr-Schild spacetimes. We extend several Kerr-Schild solutions from GR to the multigravity context. We rederive known black hole solutions in multigravity obtained using the vielbein formalism (see \cite{Wood:2024acv}) and find additional new solutions in Kerr-Schild form. Section \ref{S4} further expands the analysis of the proportional Kerr-Schild solutions from Section \ref{S3} to include double Kerr-Schild solutions, obtaining Plebański-Demiański-like solutions in the context of multigravity. After constructing solutions of the Kerr-Schild and double Kerr-Schild varieties for multigravity, Section \ref{S5} focuses on examining and extending the classical double copy in the context of multigravity where we construct spin-1 and spin-0 fields out of the interacting Kerr-Schild and double Kerr-Schild spin-2 field solutions found in multigravity. These fields referred to as single copy and zero copy fields satisfy Proca's equation and Klein-Gordon equations. In addition, we discuss the connection not only at the classical solutions level but at the Lagrangian level. In particular, we identify the multi-field gage theories underlying these maps and highlight their applications to cosmology.
Finally, in Section \ref{S6} we present the final observations of the study and discuss the next possible steps.

\vskip 1truecm
\section{Overview of Multigravity}\label{S2}
In this section, we overview the basic ideas of massive gravity, bigravity, and multigravity. We are not intended to be exhaustive, but just to provide the guide of ideas and to give the notation and conventions that we will follow in subsequent sections. 
\subsection{Massive Gravity}
General Relativity (GR) is the accepted theory that describes the gravitational field. From the perspective of particle physics, we can think of the gravitational field as a spin-2 field with massless excitations known as gravitons, which mediate the gravitational interaction. In fact, it can be shown that GR is the unique theory that can be constructed from a spin-2 massless field theory, which is local, Lorentz invariant and unitary as can be seen in \cite{Feynman:1996kb,Gupta:1954zz,Wald:1986bj,Weinberg:1965rz,Weinberg:1980kq}. One of the primary modifications to GR consists in generating a theory with a massive graviton instead of massless, this theory is commonly known as massive gravity (for a recent review, see \cite{deRham:2014zqa}). After some initial efforts in \cite{Chamseddine:2011mu,GrootNibbelink:2006vzk,Hinterbichler:2012cn,Zumino:1970tu} a particular ghost free theory of massive gravity was consistently formulated in \cite{deRham:2010kj} and it is given in terms of an action describing two interacting metrics $f$ and $g$
\begin{equation}\label{DRGT1}
    S_{mGR}=\frac{M_{Pl}^2}{2}\int d^4x \sqrt{-g}\left[R+\frac{m^2}{2}\sum_{n=0}^{4}\alpha_n \mathcal{L}_{n}[\mathcal{K}(g,f)]\right],
\end{equation}
where sgn$(g)=(-1,+1,+1,+1)$ is the signature of the metric $g$ and
\begin{equation}\label{DRGT2}
    \begin{split}
        \mathcal{L}_0(\gamma)&=1, \\
        \mathcal{L}_1(\gamma)&=[\gamma],\\
        \mathcal{L}_2(\gamma)&=\frac{1}{2!}\left([\gamma]^2-[\gamma^2]\right),\\
        \mathcal{L}_3(\gamma)&=\frac{1}{3!}\left([\gamma]^3-3[\gamma][\gamma^2]+2[\gamma]^3 \right),\\
    \mathcal{L}_4(\gamma)&=\frac{1}{4!}\left([\gamma]^4-6[\gamma^2][\gamma]^2+8[\gamma][\gamma^3]+3[\gamma^2]^2-6[\gamma^4] \right), \\
    \gamma&=\sqrt{g_k^{-1}g_{k+1}},\\
    [\gamma]&={\rm Tr}(\gamma),
    \end{split}
\end{equation}
are the symmetric polynomials in 4 dimensions, and
\begin{equation}\label{DRGT3}
    \mathcal{K}(g,f)^{\mu}_{\nu}=\delta^{\mu}_{\nu}-(\sqrt{g^{-1}f})^{\mu}_{\nu}.
\end{equation}
This theory of massive gravity is known as {\it dRGT massive gravity} due to its authors (de Rham-Gabadadze-Tolley). One of the best approaches to derive this theory is the deconstruction paradigm (see \cite{deRham:2010kj}). It was shown in \cite{Chamseddine:2011mu,deRham:2010kj,Hinterbichler:2012cn} that this theory is free of the Boulware-Deser ghost. dRGT massive gravity is a theory in which the spin-2 field named $g_{\mu \nu}$ is a dynamical tensor field which interacts with an external non-dynamical spin-2 field named $f_{\mu\nu}$. This theory of gravitation propagates 5 degrees of freedom.

\subsection{Bigravity}
As was done with massive gravity, bimetric gravity can also be derived from the deconstruction paradigm, the idea was investigated for years, as can be seen in \cite{deRham:2014zqa,Israelit:1986ez,Rosen:1978mb}. It was thought for a long time that bimetric gravity hosts a Boulware-Deser ghost until the work of Hassan and Rosen \cite{Hassan:2011zd}, where they found a consistent ghost free formulation of bimetric gravity given by the following action
\begin{equation}\label{BG1}
S_{bigravity}=\frac{1}{2\kappa_g}\int d^4x\sqrt{-g}R[g]+\frac{1}{2\kappa_f}\int d^4x\sqrt{-f}R[f]-\frac{m^2}{\kappa}\int d^4x\sqrt{-g}\sum_{n=0}^4b_n \mathcal{L}_n[\sqrt{X}],
\end{equation}
where $X=g^{-1}f$. 

In this formulation of bigravity both metrics $g$ and $f$ carry a superposition of a massless and a massive spin-2 field. The theory propagates 5+2 degrees of freedom in total. The absence of the Boulware-Deser ghost has been the subject of many discussions (see \cite{Kluson:2013jlo,Kluson:2012ps,Kluson:2013lza,Kluson:2013aca}), but they have all been clarified through rigorous proofs in \cite{Boulware:1972yco,Damour:2002wu,deRham:2010kj,Hassan:2011ea,Hassan:2011hr,Hassan:2012qv,Kluson:2012wf,Kluson:2012gz} addressing all issues.

\subsection{Multigravity}
Similarly to what was done with massive gravity and bimetric gravity, multigravity can also be derived from the deconstruction paradigm. A consistent Boulware-Deser ghost free formulation of multigravity was found in \cite{Hinterbichler:2012cn} and it is given by the following action
\begin{equation}\label{MG1}
S_N = \frac{M_{Pl}^2}{2} \sum_{j=1}^{N} \int d^4x \sqrt{-g_j}\left[R(g_j)+\frac{m_N^2}{2}\sum_{n=0}^{4}b_{n}^{j}\mathcal{L}_n(X[g_j,g_{j+1}]) \right].
\end{equation}
Multigravity is a Boulware-Deser ghost free theory of $N$ spin-2 fields interacting with first neighbors. It is a generalization of the bimetric gravity theory to $N$ interacting spin-2 fields. The theory propagates $5N-3$ degrees of freedom. Defining $N=2M+1$ the $5N-3$ degrees of freedom come from the following count: $2M$ massive spin-2 fields with five degrees of freedom each and one massless spin-2 field with 2 degrees of freedom.
The Planck mass, as presented in equation \eqref{MG1}, can be reformulated by employing the Newtonian gravitational constant corresponding to each corresponding metric
\begin{equation}\label{MG2}
    \frac{M_{Pl}^2}{2}=\frac{1}{2\kappa_j}.
\end{equation}
Henceforth, the action associated to multigravity, as delineated in equation \eqref{MG1}, which is written as follows
\begin{equation}\label{MG3}
    S_N=\sum_{j=1}^{N}\frac{1}{2\kappa_j}\int d^4x\sqrt{-g_j}\left[R(g_j)+\frac{m_N^2 M_{Pl}^2}{4}\sum_{n=0}^{4}b_{n}^{j}\mathcal{L}_n(g_j,g_{j+1})\right].
\end{equation}
Let us redefine the constants present in the interaction term in the following manner
\begin{equation}\label{MG4}
    \frac{m_N^2M_{Pl}^2}{4}=-\frac{m^2}{\kappa}.
\end{equation}
Consequently, the action can be reformulated in the form
\begin{equation}\label{mg7}
      S_N=\sum_{j=1}^{N}\int d^4x\sqrt{-g_j}\left[\frac{1}{2\kappa_j}R(g_j)-\frac{m^2}{\kappa}\mathcal{U}(g_j,g_{j+1}) \right],
\end{equation}
where the potential term ${\cal U}$ is given by
\begin{equation}\label{MG6}
    \mathcal{U}(g_j,g_{j+1})=\sum_{n=0}^{4}b_{n}^{j}\mathcal{L}_n(g_j,g_{j+1}).
\end{equation}

\subsubsection{Equations of motion}
The classical field equations of motion for multigravity can be derived by performing a variation of the action described in equation \eqref{mg7} with respect to each spin-2 field, denoted as $g_{j{\mu \nu}}$, and subsequently equating this variation to zero. Through this method, the resulting equations of motion for multigravity are written as
\begin{equation}\label{em3}
    G^{\mu}_{\nu}(g_k)=\frac{\kappa_k m^2}{\kappa}V^{\mu}_{\nu}(g_k)+\frac{\kappa_k m^2}{\kappa}W^{\mu}_{\nu}(g_k),
\end{equation}
where
\begin{equation}\label{em2}
    V_{\nu}^{\mu}=\frac{2 g^{\mu \alpha}_k}{\sqrt{-g_k}}\frac{\delta}{\delta g_k^{\alpha \nu}} \bigg( \sqrt{-g_k}\mathcal{U}(g_k,g_{k+1})\bigg),\quad W_{\nu}^{\mu}=\frac{2g_k^{\mu \alpha}}{\sqrt{-g_k}}\frac{\delta}{\delta g_k^{\alpha \nu}} \bigg( \sqrt{-g_{k-1}}\mathcal{U}(g_{k-1},g_{k})\bigg).
\end{equation}

The explicit expressions for the equations of motion associated to each spin-2 metric field are given as follows
\begin{equation}\label{em4}
    \begin{split}
        G^{\mu}_{\nu}(g_1)=&\frac{\kappa_1 m^2}{\kappa}V^{\mu}_{\nu}(g_1), \\
        G^{\mu}_{\nu}(g_k)=&\frac{\kappa_k m^2}{\kappa}V^{\mu}_{\nu}(g_k)+\frac{\kappa_k m^2}{\kappa}W^{\mu}_{\nu}(g_k),\quad k \in[2,N-1],\\
        G^{\mu}_{\nu}(g_N)=&\frac{\kappa_N m^2}{\kappa}W^{\mu}_{\nu}(g_N).        
    \end{split}
\end{equation}
Let us derive explicit expressions for each of the specified tensors present in the equations of motion as outlined in equation \eqref{em4}
\begin{equation}\label{em6}
 V^{\mu}_{\nu}(g_k)=2g_k^{\mu \alpha}\frac{\delta \mathcal{U}(g_k,g_{k+1})}{\delta g_k^{\alpha \nu}}-\delta^{\mu}_{\nu}\mathcal{U}(g_k,g_{k+1}).
\end{equation}
Analogously, the explicit expression for \( W^{\mu}_{\nu}(g_k) \) is given as follows
\begin{equation}\label{em7}
    W^{\mu}_{\nu}(g_k)=\frac{\sqrt{-g_{k-1}}}{\sqrt{-g_k}}2g_k^{\mu\alpha}\frac{\delta \mathcal{U}(g_{k-1},g_{k})}{\delta g_k^{\alpha \nu}}.
\end{equation}

Thus, the equations of motion can be rewritten in the following form
\begin{equation}\label{eom10}
    \begin{split}
        G^{\mu}_{\nu}(g_1)=&\frac{\kappa_1 m^2}{\kappa}\left[ \tau^{\mu}_{\nu}(g_1,g_2)-\mathcal{U}(g_1,g_2)\delta^{\mu}_{\nu}\right], \\
        G^{\mu}_{\nu}(g_k)=&\frac{\kappa_k m^2}{\kappa}\left[ \tau^{\mu}_{\nu}(g_k,g_{k+1})-\mathcal{U}(g_k,g_{k+1})\delta^{\mu}_{\nu}\right]-\frac{\kappa_k m^2}{\kappa}\frac{\sqrt{-g_{k-1}}}{\sqrt{-g_k}}\tau^{\mu}_{\nu}(g_{k-1},g_{k}), \\
        G^{\mu}_{\nu}(g_N)=&-\frac{\kappa_N m^2}{\kappa}\frac{\sqrt{-g_{N-1}}}{\sqrt{-g_N}}\tau^{\mu}_{\nu}(g_{N-1},g_{N}),  
    \end{split}
\end{equation}
where
\begin{equation}\label{em10}
\begin{split}
        \tau^{\mu}_{\nu}(g_k,g_{k+1})&=(b_1^{k}\mathcal{L}_0(g_k,g_{k+1})+b_2^{k}\mathcal{L}_1(g_k,g_{k+1})+b_3^{k}\mathcal{L}_2(g_k,g_{k+1}) \\ &+b_4^{k}\mathcal{L}_3(g_k,g_{k+1}))\gamma^{\mu}_{\nu}
        -(b_2^{k}\mathcal{L}_0(g_k,g_{k+1})+b_3^{k}\mathcal{L}_1(g_k,g_{k+1}) \\ &+b_4^{k}\mathcal{L}_2(g_k,g_{k+1}))(\gamma^2)^{\mu}_{\nu} +(b_3^{k}\mathcal{L}_0(g_k,g_{k+1})+b_4^{k}\mathcal{L}_1(g_k,g_{k+1}))(\gamma^3)^{\mu}_{\nu} \\
        &-b_4^{k}\mathcal{L}_0(g_k,g_{k+1})(\gamma^4)^{\mu}_{\nu}.
    \end{split}
\end{equation}

For the specific case of \( N=2 \), the equations of motion, which are deduced from equations \eqref{eom10}, can be expressed as follows
\begin{equation}\label{eom11}
    \begin{split}
        G^{\mu}_{\nu}(g)=&\frac{\kappa_g m^2}{\kappa}\left[ \tau^{\mu}_{\nu}-\mathcal{U}\delta^{\mu}_{\nu}\right], \\
        G^{\mu}_{\nu}(f)=&-\frac{\kappa_f m^2}{\kappa}\frac{\sqrt{-g}}{\sqrt{-f}}\tau^{\mu}_{\nu}.   
    \end{split}
\end{equation}
The equations presented here represent the equations of motion for bimetric gravity, as discussed in \cite{Ayon-Beato:2015qtt,Hassan:2011zd}.

\section{Proportional Kerr-Schild in Multigravity}\label{S3}

In this section we obtain several black hole solutions in multigravity using the Kerr-Schild ansatz. Some of these solutions are already reported in the literature \cite{Wood:2024acv} and  others will be new. As it was mentioned in the introduction section, for the moment we will focus in the case of proportional solutions  \cite{Hassan:2012wr, Baldacchino:2016jsz,Babichev:2015xha}. This is the case where all the metric fields are diagonal in the sense that in the description they are proportional differing among them only by a constant conformal factor. Thus the case of non-proportional solutions will be left out for further investigation.
\subsection{Equations of motion for proportional Kerr-Schild ansatz}

The initial approach in our investigation is to explore a particular type of ansatz: the proportional Kerr-Schild ansatz, which is proposed for the $N$ metric fields $g_j$. By adhering closely to the methodologies outlined in \cite{Ayon-Beato:2015qtt,Stephani:2003tm}, the formulation of the Kerr-Schild ansatz is expressed as follows
\begin{equation}\label{ks1}
    (g_j)_{\mu \nu}=C^2(j)\big[\bar{g}_{\mu \nu}+2S(j)l_{\mu}l_{\nu}\big],
\end{equation}
where $l^2=0$ and $C(1)=1$.
A fundamental characteristic of the Kerr-Schild ansatz lies in its linear dependence on the inverse metric tensors
\begin{equation}\label{ks2}
 (g_j)^{\mu \nu}=\frac{1}{C(j)^2}\big[\overline{g}^{\mu \nu}-2S(j) l^\mu l^\nu \big].
\end{equation}
As a consequence of its linearity, the $\gamma$ interaction matrix assumes a closed linear form
\begin{equation}\label{ks3}
    (\gamma(g_j,g_{j+1})^n)^{\mu}_{\nu}=\frac{C^{n}(j+1)}{C^n(j)}\bigg[\delta^{\mu}_{\nu}-n\big(S(j)-S(j+1)\big)l^{\mu}l_{\nu} \bigg].
\end{equation}
With this $\gamma$-metric the tensor $V^{\mu}_{\nu}$ can be expressed as follows
\begin{equation}\label{ks4}
    V^{\mu}_{\nu}(g_k,g_{k+1})=P_1(k)\delta^{\mu}_{\nu}-P_0(k)\big[S(k)-S(k+1)\big]l^{\mu}l_{\nu},
\end{equation}
where
\begin{equation}\label{ks5}
\begin{aligned}
P_1(k)=&\frac{\big[b_{1}(j) + 4 b_{2}(j)C(j)C(j+1) + 6 b_{3}(j)C^{2}(j)C^{2}(j+1) 
+ 4 b_{4}(j)C^{3}(j)C^{3}(j+1)\big]C(j+1)}{C(j)} \\[6pt]
&\quad - \frac{\big[b_{2}(j) + 4 b_{3}(j)C(j)C(j+1) + 6 b_{4}(j)C^{2}(j)C^{2}(j+1)\big]C^{2}(j+1)}{C^{2}(j)} \\[6pt]
&\quad + \frac{\big[b_{3}(j) + 4 b_{4}(j)C(j)C(j+1)\big]C^{3}(j+1)}{C^{3}(j)}
- \frac{b_{4}(j)C^{4}(j+1)}{C^4(j)} \\[6pt]
&\quad - b_{4}(j)C^{4}(j)C^{4}(j+1)
- 4 b_{3}(j)C^{3}(j)C^{3}(j+1)
- 6 b_{2}(j)C^{2}(j)C^{2}(j+1) \\
&\quad - 4 b_{1}(j)C(j)C(j+1)
- b_{0}(j),
\end{aligned}
\end{equation}
$$
P_0(k)=-\bigg\{
 -\frac{\big[b_{1}(k) + 4 b_{2}(k)C(k)C(k+1) + 6 b_{3}(k)C^{2}(k)C^{2}(k+1)}{C(k)}
$$
$$
+ \frac{4 b_{4}(k)C^{3}(k)C^{3}(k+1)\big]C(k+1)}{C(k)}
$$
$$
+\frac{2\big[b_{2}(k) + 4 b_{3}(k)C(k)C(k+1) + 6 b_{4}(k)C^{2}(k)C^{2}(k+1)\big]C^{2}(k+1)}{C^{2}(k)}
$$
\begin{equation}\label{ks6}
- \frac{3\big[b_{3}(k) + 4 b_{4}(k)C(k)C(k+1)\big]C^{3}(k+1)}{C^{3}(k)} 
+ \frac{4 b_{4}(k)C^{4}(k+1)}{C^{4}(k)}
\bigg\}.
\end{equation}
Moreover, the tensor $\tau^{\mu}_{\nu}$ can be represented in the form
\begin{equation}\label{ks7}
    \tau^{\mu}_{\nu}(g_k,g_{k+1})=-P_2(k)\delta^{\mu}_{\nu}-P_0(k)\big[S(k)-S(k+1)\big]l^{\mu}l_{\nu}.
\end{equation}
where
\begin{equation}\label{ks8}
\begin{aligned}
-P_2(k)=&\frac{\big[b_{1}(k) + 4 b_{2}(k)C(k)C(k+1) + 6 b_{3}(k)C^{2}(k)C^{2}(k+1) 
+ 4 b_{4}(k)C^{3}(k)C^{3}(k+1)\big]C(k+1)}{C(k)} \\[6pt]
&\quad - \frac{\big[b_{2}(k) + 4 b_{3}(k)C(k)C(k+1) + 6 b_{4}(k)C^{2}(k)C^{2}(k+1)\big]C^{2}(k+1)}{C^{2}(k)} \\[6pt]
&\quad + \frac{\big[b_{3}(k) + 4 b_{4}(k)C(k)C(k+1)\big]C^{3}(k+1)}{C^{3}(k)}
- \frac{b_{4}(k)C^{4}(k+1)}{C^{4}(k)}.
\end{aligned}
\end{equation}
Consequently, the equations of motion governing the dynamics for the Kerr-Schild ansatz can be expressed as follows
\begin{equation}\label{eq3.1.12}
    \begin{split}
    G^{\mu}_{\nu}(g_1)=&\frac{\kappa_1 m^2}{\kappa}\big[P_1(1)\delta^{\mu}_{\nu}-P_0(1)(S(1)-S(2))l^{\mu}l_{\nu} \big], \\
    G^{\mu}_{\nu}(g_k)=&\frac{\kappa_k m^2}{\kappa}\big[P_1(k)\delta^{\mu}_{\nu}-P_0(k)(S(k)-S(k+1))l^{\mu}l_{\nu}\big] \\&+\frac{\kappa_k m^2}{\kappa}\frac{C^4(k-1)}{C^4(k)}\big[P_2(k-1)\delta^{\mu}_{\nu}+P_0(k-1)(S(k-1)-S(k))l^{\mu}l_{\nu} \big], \\
    G^{\mu}_{\nu}(g_N)=&\frac{\kappa_N m^2}{\kappa}\frac{C^4(N-1)}{C^4(N)}\big[P_2(N-1)\delta^{\mu}_{\nu}+P_0(N-1)(S(N-1)-S(N))l^{\mu}l_{\nu} \big].
    \end{split}
\end{equation}

\subsection{A first exploration: Proportional Kerr solution}
In this subsection, we study the dynamics of Kerr-Schild ansatz given in equation \eqref{eq3.1.12} by considering metrics exhibiting axial symmetry
\begin{equation}\label{eq3.1.1}
    ds_k^2=(g_k)_{ \mu\nu}dx^{\mu}dx^{\nu}=C^2(k)\bigg[ds_M^2 +2S(k) l \otimes l\bigg],
\end{equation}
where $C(1)=1$ and
\begin{equation}\label{eq3.1.2}
    ds^2_M=-dt^2+(r^2+a^2)\sin^2\theta d\phi^2+\frac{\Sigma}{r^2+a^2}dr^2+\Sigma d\theta^2,
\end{equation}
\begin{equation}\label{eq3.1.3}
    l=-dt-a\sin^2\theta d\phi+\frac{\Sigma}{r^2+a^2}dr
\end{equation}
and
\begin{equation}\label{eq3.1.4}
    S(k)=S_k(r,\theta),\quad C(k)=C_k.
\end{equation}
\subsubsection{Circularity Theorem}
We formulate a theorem of circularity applicable to the Kerr-Schild ansatz in the context of multigravity, with the objective of deriving a multi-Kerr solution. This will be achieved by closely following and extending the methodology previously applied to the bigravity scenario as documented in \cite{Ayon-Beato:2015qtt}. Initially, we consider a stationary axial symmetric spacetime characterized by commuting Killing vector fields $k=\partial_t$ and $m=\partial_{\phi}$. In such a spacetime, the following conditions hold
\begin{equation}\label{kct1}
\star(k\wedge m \wedge dk)_k=-\frac{2l_t C_k^2 \sin\theta }{\Sigma}\frac{\partial}{\partial \theta}\left(\Sigma S_k \right),
\end{equation}
\begin{equation}\label{kct2}
\star(k\wedge m \wedge dm)_k=\frac{2l_t C_k^2 a\sin^3\theta }{\Sigma}\frac{\partial}{\partial \theta}\left(\Sigma S_k \right),
\end{equation}
where $\star$ is the Hodge star operation in the four-dimensional space. Then we have the following relations
\begin{equation}\label{kct3}
    \frac{\star(k\wedge m \wedge dk)_k}{l_t}=\frac{\star(k\wedge m \wedge dm)_k}{l_{\phi}}=-\frac{2l_t C_k^2 \sin\theta }{\Sigma}\frac{\partial}{\partial \theta}\left(\Sigma S_k \right).
\end{equation}
In addition, we have
\begin{equation}\label{kct4}
    \frac{\star(k\wedge m\wedge V(k))_k}{l_t}=-P_0(k)(S_k-S_{k+1})\star(k \wedge m \wedge l),
\end{equation}
\begin{equation}\label{kct5}
    \frac{\star(k\wedge m\wedge V(m))_k}{l_{\phi}}=-P_0(k)(S_k-S_{k+1})\star(k \wedge m \wedge l),
\end{equation}
\begin{equation}\label{kct6}
    \frac{\star(k\wedge m\wedge \tau(k))_k}{l_t}=-P_0(k-1)(S_k-S_{k-1})\star(k \wedge m \wedge l),
\end{equation}
\begin{equation}\label{kct7}
    \frac{\star(k\wedge m\wedge \tau(m))_k}{l_{\phi}}=-P_0(k-1)(S_k-S_{k-1})\star(k \wedge m \wedge l).
\end{equation}
For any stationary axisymmetric spacetime with commuting Killing vector fields $k=\partial_t$ and $m=\partial_{\phi}$, the following geometric identities hold
\begin{equation}\label{kct8}
    \begin{split}
        I_k=&d\star(k\wedge m \wedge dk)_k-2\star(k\wedge m\wedge R(k))_k=0,\\
        I_m=&d\star(k \wedge m \wedge dm)_k -2 \star(k\wedge m \wedge R(m))_k=0,
    \end{split}
\end{equation}
where
\begin{equation}\label{kct9}
    R(k)_k =R_{\mu \nu}(g_k)k^{\nu}dx^{\mu}, \quad R(m)_k=R_{\mu \nu}(g_k)m^{\nu}dx^{\mu}.
\end{equation}
Using the relationships provided in equation \eqref{kct8}, we derive the relation
\begin{equation}\label{kct10}
    \frac{l_{\phi}}{l_t}I_k-I_m=-\star(k \wedge m \wedge dk)d\left(\frac{l_{\phi}}{l_t} \right)=0.
\end{equation}
Given that the derivatives of \( l_i \) are non-zero, the only method to fulfill the identity of equation \eqref{kct10} is given by
\begin{equation}\label{kct11}
    k \wedge m \wedge dk=k \wedge m \wedge dm=0.
\end{equation}
Equation \eqref{kct11} is known as the circularity condition. By incorporating this condition into the equation \eqref{kct3}, the equation is simplified in the form
    \begin{equation}\label{kct12}
\star(k\wedge m \wedge dk)_k=-\frac{2l_t C_k^2 \sin\theta }{\Sigma}\frac{\partial}{\partial \theta}\left(\Sigma S_k \right)=0.
\end{equation}
The solution of equation \eqref{kct12} can be written as
\begin{equation}\label{kct13}
    S_k(r,\theta)=\frac{r M_k(r)}{\Sigma}.
\end{equation}
By incorporating the circularity condition into the equation \eqref{kct8} for $I_k$ and conducting an iterative resolution for each spin-2 field, we obtain the following constraint
\begin{equation}\label{kct14}
    P_0(k)=0.
\end{equation}
We implement the constraint given by Eq. \eqref{kct13} into the equations of motion, the system is consequently simplified, and this leads to
$$
G^{\mu}_{\nu}(g_1)=\frac{\kappa_1 m^2}{\kappa}\big[P_1(1)\delta^{\mu}_{\nu} \big], 
$$
$$
G^{\mu}_{\nu}(g_k)=\frac{\kappa_k m^2}{\kappa}\big[P_1(k)\delta^{\mu}_{\nu}\big] +\frac{\kappa_k m^2}{\kappa}\frac{C^4(k-1)}{C^4(k)}\big[P_2(k-1)\delta^{\mu}_{\nu}\big],
$$
\begin{equation}\label{kct15}
G^{\mu}_{\nu}(g_N)=\frac{\kappa_N m^2}{\kappa}\frac{C^4(N-1)}{C^4(N)}\big[P_2(N-1)\delta^{\mu}_{\nu}\big].
\end{equation}
The explicit computation of the $G^{r}_{r}(g_k)$ component of the Einstein tensor yields the following equation
\begin{equation}\label{kct16}
    G^{r}_r(g_k)=-\frac{2M_k'(r)}{\Sigma^2 C(k)^2}=\frac{\kappa_k m^2}{\kappa}\left(P_1(k)+\frac{C^4(k-1)}{C^4(k)}P_2(N-1) \right).
\end{equation}
The only way to fulfill equation \eqref{kct16} is
\begin{equation}\label{kct17}
\begin{split}
M_1(r)=&m_1,\quad \ \ P_1(1)=0,\\
    M_k(r)=&m_k,\quad \ \ P_1(k)+\frac{C^4(k-1)}{C^4(k)}P_2(k-1)=0,\\
    M_N(r)=&m_N,\quad \ \ P_2(N-1)=0.
\end{split}
\end{equation}
Consequently, the axisymmetric Kerr-Schild solution within multigravity is expressed as
$$
ds_k^2=C^2(k)\bigg[ds_M^2+\frac{2m_kr}{\Sigma} l \otimes l\bigg],
$$   
\begin{equation}\label{kct18}
\begin{split}   
    P_0(k)&=0, \quad 1\leq k \leq N,\\
    P_1(1)&=0,\\
     P_1(k)+\frac{C^4(k-1)}{C^4(k)}P_2(k-1)&=0,\quad 2\leq k \leq N-1,\\
     P_2(N-1)&=0.
\end{split}
\end{equation}
Equation \eqref{kct18} is an axial-symmetric solution of multigravity, characterized by a sequence of $N$ proportional Kerr black holes. These black holes are distinguished by possessing identical angular momenta while maintaining independent distinct mass values. In the scenario where \( N=2 \), the solution reduces to the bimetric Kerr solution of bigravity found in \cite{Ayon-Beato:2015qtt}. This solution was obtained as a limiting case of the Kerr-Newman-AdS metric found by Wood et al in \cite{Wood:2024acv} using the vielbein formalism, and here we obtained it in the metric formulation.

\subsection{A first exploration: Proportional Kerr-AdS solution }
Now we proceed with our exploration by examining an axially symmetric proportional Kerr-Schild ansatz within an AdS maximally symmetric background, as described below
\begin{equation}\label{kads1}
    ds^2_k=C^2(k)\left(ds_0^2+{2S(k)}l \otimes l \right),
\end{equation}
where
\begin{equation}\label{kads2}
    ds_0^2=-\frac{(1-\lambda r^2)\Delta_{\theta}}{1+\lambda a^2}dt^2+\frac{(r^2+a^2)\sin^2\theta}{1+\lambda a^2}d\phi^2+\frac{\Sigma}{(1-\lambda r^2)(r^2+a^2)}dr^2+\frac{\Sigma}{\Delta_{\theta}}d\theta,
\end{equation}
\begin{equation}\label{kads3}
    l=\frac{\Delta_{\theta}}{1+\lambda a^2}dt-\frac{a\sin^2 \theta}{1+\lambda a^2}d\phi +\frac{\Sigma}{(1-\lambda r^2)(r^2+a^2)}dr
\end{equation}
and $\Delta_{\theta}=1+\lambda a^2 \cos^2\theta$.
\subsubsection{Circularity theorem}
In stationary axisymmetric spacetimes, such as \eqref{kads1}, there are two commuting Killing vector fields: $k=\partial_t$ and $m=\partial_{\phi}$. Within these spacetimes, the following conditions are satisfied
\begin{equation}\label{kads4}
    \frac{\star(k\wedge m \wedge dk)}{l_t}=\frac{\star(k \wedge m \wedge dm)}{l_{\phi}}=-\frac{2C_k^2 \sin\theta \Delta_{\theta}}{\Sigma (1+\lambda a^2)} \frac{\partial}{\partial \theta}(\Sigma S_k).
\end{equation}
In addition, we have
\begin{equation}\label{kads5}
    \frac{\star(k\wedge m\wedge V(k))_k}{l_t}=-\frac{P_0(k) \Delta_{\theta}}{1+\lambda a^2}(S_k-S_{k+1})\star(k \wedge m \wedge l),
\end{equation}\label{kads6}
\begin{equation}
    \frac{\star(k\wedge m\wedge V(m))_k}{l_{\phi}}=-\frac{P_0(k) \Delta_{\theta}}{1+\lambda a^2}(S_k-S_{k+1})\star(k \wedge m \wedge l),
\end{equation}

\begin{equation}\label{kads7}
    \frac{\star(k\wedge m\wedge \tau(k))_k}{l_t}=-\frac{P_0(k-1) \Delta_{\theta}}{1+\lambda a^2}(S_k-S_{k-1})\star(k \wedge m \wedge l),
\end{equation}
\begin{equation}\label{kads8}
    \frac{\star(k\wedge m\wedge \tau(m))_k}{l_{\phi}}=-\frac{P_0(k-1) \Delta_{\theta}}{1+\lambda a^2}(S_k-S_{k-1})\star(k \wedge m \wedge l).
\end{equation}
By substituting the conditions delineated in equations \eqref{kads4} through \eqref{kads8} into equation \eqref{kct8}, we derive the following condition
\begin{equation}\label{kads9}
    \frac{l_{\phi}}{l_t}I_k-I_m=-\star(k \wedge m \wedge dk)d\left(\frac{l_{\phi}}{l_t} \right)=0.
\end{equation}
Given that the derivatives of \( l_i \) are non-zero, the only way to fulfill the identity is that
\begin{equation}\label{kads10}
    k \wedge m \wedge dk=k \wedge m \wedge dm=0.
\end{equation}
Equation \eqref{kads10} is known as the circularity condition. By incorporating this condition into the equation \eqref{kads4}, the equation is simplified as follows: 
\begin{equation}\label{kads11}
\star(k\wedge m \wedge dk)_k=-\frac{2C_k^2 \sin\theta \Delta_{\theta}}{\Sigma (1+\lambda a^2)} \frac{\partial}{\partial \theta}(\Sigma S_k)=0.
\end{equation}
The solution to equation \eqref{kads11} is subsequently determined in the following form
\begin{equation}\label{kads12}
    S_k(r,\theta)=\frac{r M_k(r)}{\Sigma}.
\end{equation}
The implementation of the circularity condition in $I_k$ and performing an iterative resolution for each spin-2 field, the following constraints are obtained
\begin{equation}\label{kads13}
    P_0(k)=0.
\end{equation}
By incorporating the constraint given in equation \eqref{kads12} into equations of motion, the system of equations is consequently simplified in the form
$$
G^{\mu}_{\nu}(g_1)=\frac{\kappa_1 m^2}{\kappa}\big[P_1(1)\delta^{\mu}_{\nu} \big], 
$$
$$
G^{\mu}_{\nu}(g_k)=\frac{\kappa_k m^2}{\kappa}\big[P_1(k)\delta^{\mu}_{\nu}\big] +\frac{\kappa_k m^2}{\kappa}\frac{C^4(k-1)}{C^4(k)}\big[P_2(k-1)\delta^{\mu}_{\nu}\big],
$$
\begin{equation}\label{kads14}
    G^{\mu}_{\nu}(g_N)=\frac{\kappa_N m^2}{\kappa}\frac{C^4(N-1)}{C^4(N)}\big[P_2(N-1)\delta^{\mu}_{\nu}\big].
\end{equation}
The only way to fulfill equation \eqref{kads14} is
\begin{equation}\label{kads15}
\begin{split}
M_1(r)&=m_1,\quad P_1(1)=-\frac{3 \lambda \kappa}{\kappa_k m^2 C(1)^2},\\
    M_k(r)&=m_k,\quad P_1(k)+\frac{C^4(k-1)}{C^4(k)}P_2(k-1)=-\frac{3 \lambda \kappa}{\kappa_k m^2 C^2(k)},\\
    M_N(r)&=m_n,\quad \frac{C^4(N-1)}{C^4(N)}P_2(N-1)=-\frac{3 \lambda \kappa}{\kappa_k m^2 C^2(N)}.
\end{split}
\end{equation}
Consequently, the axisymmetric Kerr-Schild solution in an AdS background within multigravity is expressed as follows
\begin{equation}\label{kads16}
\begin{split}
    ds_k^2=&C(k)^2\bigg[ds_0^2+\frac{2m_kr}{\Sigma} l \otimes l  \bigg],\\
        P_0(k)&=0, \quad 1\leq k \leq N,\\
    P_1(1)&=-\frac{3 \lambda \kappa}{\kappa_1 m^2 C^2(k)},\\
     P_1(k)+\frac{C^4(k-1)}{C^4(k)}P_2(k-1)&=-\frac{3 \lambda \kappa}{\kappa_k m^2 C^2(k)},\quad 2\leq k \leq N-1,\\
     \frac{C^4(N-1)}{C^4(N)}P_2(N-1)&=-\frac{3 \lambda \kappa}{\kappa_N m^2 C^2(N)}.
\end{split}
\end{equation}
Equation \eqref{kads16} is an axial-symmetric solution in an AdS background of multigravity, characterized by a sequence of $N$ proportional Kerr-AdS black holes. These black holes are distinguished by possessing identical angular momenta and being in the same AdS background while maintaining distinct and independent mass values. In the scenario where $N=2$, the solution reduces to the Kerr-AdS bimetric solution of bigravity found in \cite{Ayon-Beato:2015qtt}. This solution was obtained as a limiting case of the Kerr-Newman-AdS metric found in \cite{Wood:2024acv} in the vielbein formulation, and we rederived it here in the metric formulation.
\subsection{Proportional Kerr-Schild solutions in vacuum }
In addition to the Multi-Kerr and Multi-Kerr-AdS solutions identified in this paper, we should examine what other solutions in the proportional Kerr-Schild form might be discovered by scrutinizing the equations of motion for the proposed proportional Kerr-Schild solutions referred to in equation \eqref{eq3.1.12}. According to the contracted Bianchi identity, it is known that the Einstein tensor is covariantly conserved in the following manner
\begin{equation}\label{gks1}
    \nabla_{\mu}G^{\mu}_{\nu}(g_k)=0.
\end{equation}
By applying the covariant derivative to the equations of motion for multigravity concerning the proportional Kerr-Schild solutions specified in \eqref{eq3.1.12}, we get the following constraints
$$
P_0(1)\nabla_{\mu}\bigg[\big(S(1)-S(2)\big)l^{\mu}l_{\nu} \bigg]=0,
$$
$$      
-P_0(k)\nabla_{\mu} \bigg[\big(S(k)-S(k+1)\big)l^{\mu}l_{\nu}\bigg]+\frac{C^4(k-1)}{C^4(k)}P_0(k-1)\big[S(k-1)-S(k) \big]=0, 
$$
\begin{equation}\label{gkos2}
P_0(N-1)\big[S(N-1)-S(N) \big]=0.
\end{equation}
The constraints given in equation \eqref{gkos2} are satisfied if, and only if
\begin{equation}\label{gks3}
\begin{split}
    P_0(k)=0, 
\end{split}
\end{equation}
for all values of $k$ in $1\leq k < N$.
Upon integrating out the results derived from equation \eqref{gks3} into the equations of motion as delineated for the proportional Kerr-Schild proposed solutions in equation \eqref{eq3.1.12}, the resultant equations simplify as follows
$$
G^{\mu}_{\nu}(g_1)=\frac{\kappa_1 m^2}{\kappa}\big[P_1(1)\delta^{\mu}_{\nu}\big],
$$
$$ 
G^{\mu}_{\nu}(g_k)=\frac{\kappa_k m^2}{\kappa}\bigg[P_1(k)\delta^{\mu}_{\nu}\bigg] +\frac{\kappa_k m^2}{\kappa}\frac{C^4(k-1)}{C^4(k)}\bigg[P_2(k-1)\delta^{\mu}_{\nu} \bigg],
$$
\begin{equation}\label{gks4}
G^{\mu}_{\nu}(g_N)=\frac{\kappa_N m^2}{\kappa}\left[\frac{C^4(N-1)}{C^4(N)}P_2(N-1)\delta^{\mu}_{\nu} \right].
\end{equation}
In order to proceed with the computation of the Ricci scalar of curvature we contract the free indices in equation \eqref{gks4} and find the following
\begin{equation}\label{gks5}
    \begin{split}
        R(g_1)=&-\frac{\kappa_1 m^2}{\kappa}\bigg[4P_1(1)\bigg], \\
        R(g_k)=& -\frac{\kappa_k m^2}{\kappa}\left[4P_1(k)+4\frac{C^4(k-1)}{C^4(k)} P_2(k-1)\right], \\
        R(g_N)=&-\frac{\kappa_N m^2}{\kappa}\left[4\frac{C^4(N-1)}{C^4(N)}P_2(N-1) \right].
    \end{split}
\end{equation}
However, the functions \(P_1(k)\) and \(P_2(k)\), as specified in equations \eqref{ks5} and \eqref{ks8}, are constant scalars, as are the gravitational constant corresponding to each spin-2 field metric, as well as the interaction parameters \(\kappa_k,\kappa, m^2\). Consequently, the right-hand side of the equation \eqref{gks5} represents a constant scalar. Therefore, in the theory of multigravity, the only viable solutions derivable from a proportional Kerr-Schild ansatz are spin-2 metric fields characterized by a constant Ricci scalar curvature.

Consider the definition of the parameters denoted as $\Lambda_k$:
\begin{equation}\label{gks6}
\begin{split}
    \Lambda_1=&-\frac{\kappa_1m^2}{\kappa}P_1(1),\\
    \Lambda_k=&-\frac{\kappa_k m^2}{\kappa}\left(P_1(k)+\frac{C^4(k-1)}{C^4(k)}P_2(k-1) \right), \quad k \in[2,N-1],\\
    \Lambda_N =&-\frac{\kappa_Nm^2}{\kappa}\frac{C^4(N-1)}{C^4(N)}P_2(N-1),
\end{split}
\end{equation}

By substituting the parameters denoted as $\Lambda_k$ into the equations of motion as presented in equation \eqref{gks4}, and subsequently performing a contraction with the term $(g_k)_{\mu \alpha}$, the equations of motion for a proportional Kerr-Schild ansatz in multigravity are reformulated in the following form
\begin{equation}\label{gks7}
G_{\mu \nu}(g_k)+\Lambda_k (g_k)_{ \mu \nu}=0, 
\end{equation}
where $k \in[1,N]$.
This implies that any Kerr-Schild solution within GR, characterized by a constant Ricci scalar, can serve as a foundational basis to formulate a proportional Kerr-Schild solution in the context of multigravity. In this scenario, the Ricci scalar curvatures $R(g_k)$, cosmological constants $\Lambda_k$, and the free parameters of the metrics $C(k)$ are not mutually independent but are subject to constraints imposed by the equations \eqref{gks5} and \eqref{gks6}. In addition, for this specific type of solutions, a cosmological constant term emerges in the equations of motion \eqref{gks7} due to the interaction potential between gravitons $\mathcal{U}$ as specified in \eqref{mg7}. It is worth noticing that even when the cosmological constant interaction term $\mathcal{L}_0$ in $\mathcal{U}$ is set to zero, the cosmological constant appearing in \eqref{gks7} remains. Thus for Kerr-Schild solutions the cosmological constants $\Lambda_k$ have an easy intuitive physical explanation, they are energy densities generated by the interaction potential among different kinds of gravitons $\mathcal{U}$.

Consequently, solutions to multigravity can be constructed in the following manner. One can propose proportional Kerr-Schild spacetimes as follows
\begin{equation}\label{ksextra1}
    \begin{split}
        (g_j)_{\mu \nu}=C^2(j)\big[\overline{g}_{\mu \nu}+2S_jk_{\mu}k_{\nu} \big].
    \end{split}
\end{equation}
In this context, $\overline{g}_{\mu \nu}$ represents the metric tensor associated with a maximally symmetric spacetime, while $(g_k)_{\mu \nu}$ denotes a Kerr-Schild spacetime characterized by a constant Ricci curvature that satisfies Einstein's field equations
$$
G_{\mu \nu}(g_k)+\Lambda_k (g_k)_{ \mu \nu}=0, \ \ \ \  \forall \ k \in[1,N],
$$
$$
 P_0(k)=0,
$$
\begin{equation}\label{ksextra}
\begin{split}     
    \Lambda_1=&-\frac{\kappa_1m^2}{\kappa}P_1(1),\\
    \Lambda_k=&-\frac{\kappa_k m^2}{\kappa}\left(P_1(k)+\frac{C^4(k-1)}{C^4(k)}P_2(k-1) \right),\quad \forall \ k \in[2,N-1],\\
    \Lambda_N =&-\frac{\kappa_Nm^2}{\kappa}\frac{C^4(N-1)}{C^4(N)}P_2(N-1).
\end{split}
\end{equation}
In the next subsections, we will focus on constructing solutions of multigravity by employing the methodology of proportional Kerr-Schild spacetimes specified previously.
\subsubsection{Multi-Schwarzschild}
To construct multigravity vacuum solutions from proportional Kerr-Schild spacetimes, we take the Schwarzschild metric in Kerr-Schild form as a seed and follow the prescription in \eqref{ksextra}. In this representation, the Schwarzschild metric reads as follows
\begin{equation}\label{sch1}
    g_{\mu \nu} = \eta_{\mu \nu}+\frac{2GM}{r}k_{\mu}k_{\nu},
\end{equation}
with $k_{\mu}=(1,\widehat{r})$. The metric \eqref{sch1} satisfies the vacuum Einstein equations $G_{\mu \nu}=0$, and describes a static, spherically symmetric spacetime of constant curvature, with scalar curvature $R=0$. Therefore, by using this spin-2 field as input and applying the procedure of \eqref{ksextra}, we obtain the corresponding spherically symmetric multigravity configuration
$$
g_{\mu \nu}(j)= C^2({j})\left[\eta_{\mu \nu}+\frac{2G_j m_j}{r}k_{\mu}k_\nu \right],
$$
\begin{equation}\label{sch2}
\begin{split}
    P_0(j)=&0,\quad 1\leq j \leq N, \\
    P_1(1) =&0, \\
    P_1(j)+\frac{C^4(j-1)}{C^4(j)}P_2(j-1)=&0, \quad 2 \leq j \leq N-1,\\
    P_2(N-1)=&0.
\end{split}
\end{equation}
Equation \eqref{sch2} yields a spherically symmetric, static vacuum solution in multigravity. In particular, it describes a chain of $N$ Schwarzschild black holes, with the masses $m_j$ taken as independent parameters for each sector. As in related constructions discussed above, this configuration can be recovered as a limiting case of the Kerr-Newman-AdS metric of Ref. \cite{Wood:2024acv}, and it is rederived here within the metric formulation. 
\subsubsection{Multi-Schwarzschild-AdS}
As a natural continuation of the Schwarzschild configuration in \eqref{sch2}, we now incorporate a negative cosmological constant by choosing the Schwarzschild Anti-de Sitter (AdS) metric in Kerr-Schild form as the seed metric. In this representation we have
\begin{equation}\label{sch3}
    g_{\mu \nu} = \overline{g}_{\mu \nu}+\frac{\kappa}{2}\frac{m}{4\pi r}k_{\mu}k_{\nu},
\end{equation}
where $\overline{g}_{\mu \nu}$ denotes the AdS background metric
\begin{equation}\label{sch4}
    \overline{g}_{\mu \nu} ={\rm diag} \left(1-\frac{\Lambda r^2}{3}, \left(1-\frac{\Lambda r^2}{3}\right)^{-1},r^2,r^2 \sin^2\theta\right)
\end{equation}
and the associated Kerr-Schild null vector is
\begin{equation}\label{sch5}
    k_{\mu}=\left(1,\frac{1}{1-\frac{\Lambda r^2}{3}},0,0\right).
\end{equation}
Starting from this Schwarzschild-AdS input, the multigravity counterpart follows directly from the proportional Kerr-Schild construction and can be written as
$$
g_{\mu \nu}(j)=C^2(j)\left[\overline{g}_{\mu \nu}+\frac{\kappa_j m_j}{8\pi r}k_{\mu} k_{\nu}\right],
$$
$$
P_0(j)=0, \ \ \ \ \ 1\leq j \leq N,
$$
\begin{equation}\label{sch6}
    \begin{split}
    \Lambda=&-\frac{\kappa_1m^2}{\kappa}P_1(1),\\
    \Lambda=&-\frac{\kappa_k m^2}{\kappa}\left(P_1(j)+\frac{C^4(j-1)}{C^4(j)}P_2(j-1) \right), \quad  2\leq j \leq N-1,\\
    \Lambda =&-\frac{\kappa_Nm^2}{\kappa}\frac{C^4(N-1)}{C^4(N)}P_2(N-1).    \end{split}
\end{equation}
Equation \eqref{sch6} defines the AdS extension of the previous vacuum construction: it corresponds to a collection of $N$ Schwarzschild black holes with independent masses, now embedded in a common Anti-de Sitter (AdS) background. In particular, all sectors share the same cosmological constant $\Lambda$. As before, the solution may be obtained as a suitable limiting case of the Kerr-Newman-AdS family in \cite{Wood:2024acv} (derived there in the vielbein formalism), and we rederive it here in the metric formulation.

\subsubsection{Multi-Kundt waves}
Next, we consider Kundt wave solutions on an Anti-de Sitter (AdS) background, which admit a convenient Kerr-Schild description. In this case, the background metric takes the form \cite{Carrillo-Gonzalez:2017iyj}
$$
\overline{g}_{\mu \nu}dx^{\mu}dx^{\nu}=\frac{1}{P^2}\big[-4x^2du(dv-v^2du)+dx^2+dy^2 \big],
$$
\begin{equation}\label{kundt1}
P=1+\frac{\Lambda}{12}(x^2+y^2),
\end{equation}
where $u,v$ are the light-cone coordinates. With this choice, the full Kerr-Schild metric is written as
\begin{equation}\label{kundt2}
    g_{\mu \nu} = \overline{g}_{\mu \nu}+\frac{P}{x}H(u,x,y)k_{\mu}k_{\nu},
\end{equation}
with
\begin{equation}
    k_{\mu}=\frac{x}{P}\delta^{\mu}_{\mu}.
\end{equation}
Moreover, the profile function \( H(u,x,y) \) is constrained by the equation of motion
\begin{equation}\label{kundth}
    \left(\partial_x^2+\partial_y^2+\frac{2\Lambda}{3P^2} \right)H(u,x,y)=0.
\end{equation}
Having \eqref{kundt2} as our input metric, the corresponding multigravity configuration can then be assembled as follows
$$
g_{\mu \nu}(j)= C^2(j)\left[ \overline{g}_{\mu \nu}+\frac{P}{x}H_j(u,x,y)k_{\mu}k_{\nu}\right],
$$
$$
P_0(j)=0 \ \ \ \ \ \forall \ j \in [1,N],
$$
\begin{equation}\label{kundtf}
    \begin{split}    
     \Lambda=&-\frac{\kappa_1m^2}{\kappa}P_1(1),\\
    \Lambda=&-\frac{\kappa_k m^2}{\kappa}\left(P_1(j)+\frac{C^4(j-1)}{C^4(j)}P_2(j-1) \right), \quad \forall \ j \in[2,N-1],\\
    \Lambda =&-\frac{\kappa_Nm^2}{\kappa}\frac{C^4(N-1)}{C^4(N)}P_2(N-1).   
    \end{split}
\end{equation}
Equation \eqref{kundtf} is a multigravity solution built consisting of $N$ Kundt-waves on a single Anti-de Sitter (AdS) background. The cosmological constant is common to all sectors, whereas the wave profiles may vary from one metric to another through the functions $H_j$.
\subsubsection{Multi-$pp$-waves}
We now turn to the generalized Anti-de Sitter (AdS) plane-fronted waves with parallel rays ($pp$-waves), which provide vacuum solutions of GR. In Kerr-Schild form they are naturally described in terms of an AdS background metric, as discussed in \cite{Carrillo-Gonzalez:2017iyj}
\begin{equation}\label{pp1}
    \overline{g}_{\mu \nu}dx^{\mu}dx^{\nu}=\frac{1}{P^2}\left[ -2Q^2\left(dv-\frac{\Lambda}{6}v^2du\right)+dx^2+dy^2\right],
\end{equation}
where
\begin{equation}\label{pp2}
    P=1+\frac{\Lambda}{12}(x^2+y^2), \ \ \ \ \ Q=1-\frac{\Lambda}{12}(x^2+y^2).
\end{equation}
Then the full metric is given by
\begin{equation}\label{pp3}
    g_{\mu \nu}=\overline{g}_{\mu \nu}+e^{2\tanh^{-1}(P-1)}H(u,x,y) k_{\mu}k_{\nu},
\end{equation}
where $H(u,x,y)$ satisfies Eq. \eqref{kundth} and
\begin{equation}\label{pp4}
    k_{\mu}=e^{-\tanh^{-1}(P-1)} \sqrt{\frac{Q}{P}}\delta^{\mu}_{\mu}.
\end{equation}
Following the same construction as in the previous examples, the associated multigravity solution can be written as
$$
g_{\mu \nu}(j)=C^2(j)\left[ \overline{g}_{\mu \nu}+e^{2\tanh^{-1}(P-1)}H(u,x,y) k_{\mu}k_{\nu}\right],
$$
$$
    P_0(j)=0,\quad \forall \ j \in [1,N],
$$
\begin{equation}\label{pp5}
    \begin{split}
     \Lambda=&-\frac{\kappa_1m^2}{\kappa}P_1(1),\\
    \Lambda=&-\frac{\kappa_k m^2}{\kappa}\left(P_1(j)+\frac{C^4(k-1)}{C^4(j)}P_2(j-1) \right),\ \ \ \ \forall  \ j \in [2,N-1],\\
    \Lambda =&-\frac{\kappa_Nm^2}{\kappa}\frac{C^4(N-1)}{C^4(N)}P_2(N-1).   
    \end{split}
\end{equation}
Equation \eqref{pp5} then gives a multigravity solution consisting of $N$ $pp$-wave solutions. All metrics propagate on the same AdS background (hence sharing the same cosmological constant), while the individual wave amplitudes are allowed to differ and are encoded in the functions $H_j$.
\subsubsection{Multi-Siklos-AdS waves}
In a similar spirit, the Siklos anti-de Sitter (AdS) waves \cite{siklos1985lobatchevski,Ayon-Beato:2018hxz} furnish a vacuum solution to Einstein's field equations and can also be cast in Kerr-Schild form, provided we adopt an appropriate choice of background metric
\begin{equation}\label{siklos1}
    \overline{g}_{\mu \nu}dx^{\mu}dx^{\nu}=\frac{l^2}{x^2}\bigg[-2dudv+dx^2+dy^2 \bigg].
\end{equation}
With this background, the full metric takes the Kerr-Schild form
\begin{equation}\label{siklos2}
    g_{\mu \nu}=\overline{g}_{\mu \nu}+\frac{x}{l}H(u,x,y)k_{\mu}k_{\nu},
\end{equation}
where
\begin{equation}\label{siklos3}
    k_{\mu}=\frac{l}{x}\delta^{\mu}_{\mu}
\end{equation}
and the profile $H(u,x,y)$ obeys Eq. \eqref{kundth}. Building on this seed solution, the multigravity generalization is obtained in the following way
$$
g_{\mu \nu}(j)=C^2(j)\bigg[\overline{g}_{\mu \nu}+\frac{x}{l}H_j(u,x,y)k_{\mu}k_{\nu}\bigg],
$$
$$
    P_0(j)=0,\ \ \  \forall \ j \in [1,N],
$$
\begin{equation}\label{siklos4}
    \begin{split}   
     \Lambda=&-\frac{\kappa_1m^2}{\kappa}P_1(1),\\
    \Lambda=&-\frac{\kappa_k m^2}{\kappa}\left(P_1(j)+\frac{C^4(k-1)}{C^4(j)}P_2(j-1) \right),\ \ \forall \ j \in [2,N-1],\\
    \Lambda =&-\frac{\kappa_Nm^2}{\kappa}\frac{C^4(N-1)}{C^4(N)}P_2(N-1).  
    \end{split}
\end{equation}
Equation \eqref{siklos4} yields a multigravity vacuum solution described by $N$ Siklos--anti-de Sitter (AdS) waves. As in the previous solutions based on AdS, the cosmological constant is the same in all sectors, whereas the wave profiles are sector-dependent and are specified by functions $H_j$.
\subsection{Proportional Kerr-Schild solutions coupled to matter}
According to the results presented in \cite{deRham:2010kj}, in multigravity theories, matter couplings that remain free from Boulware-Deser ghosts are those in which matter fields of type $k$ interact exclusively with a single spin-2 field of type $k$, as well as with other matter fields of type $k$. Consequently, within the framework of the Kerr-Schild ansatz, the equations of motion in multigravity are modified by incorporating an energy-momentum tensor $T^{\mu}_{\nu}(g_k)$ for each metric $g_k$ in the form
\begin{equation}\label{kss1}
    \begin{split}
    G^{\mu}_{\nu}(g_1)=&\kappa_1 T^{\mu}_{\nu}(g_1)+\frac{\kappa_1 m^2}{\kappa}\big[P_1(1)\delta^{\mu}_{\nu}-P_0(1)(S(1)-S(2))l^{\mu}l_{\nu} \big], \\
    G^{\mu}_{\nu}(g_k)=&\kappa_k T^{\mu}_{\nu}(g_k)+\frac{\kappa_k m^2}{\kappa}\big[P_1(k)\delta^{\mu}_{\nu}-P_0(k)(S(k)-S(k+1))l^{\mu}l_{\nu}\big] \\&+\frac{\kappa_k m^2}{\kappa}\frac{C^4(k-1)}{C^4(k)}\big[P_2(k-1)\delta^{\mu}_{\nu}+P_0(k-1)(S(k-1)-S(k))l^{\mu}l_{\nu} \big], \\
    G^{\mu}_{\nu}(g_N)=&\kappa_N T^{\mu}_{\nu}(g_N)+\frac{\kappa_N m^2}{\kappa}\frac{C^4(N-1)}{C^4(N)}\big[P_2(N-1)\delta^{\mu}_{\nu}+P_0(N-1)(S(N-1)-S(N))l^{\mu}l_{\nu} \big].
    \end{split}
\end{equation}
To initiate the examination of these inhomogeneous equations of motion, we perform a contraction of equation \eqref{kss1} with the covariant derivative pertinent to each spin-2 metric field $(g_k)_{\mu \nu}$. We take into account the covariant conservation of the Einstein tensors, as dictated by the contracted Bianchi identities $\nabla_{\mu}G^{\mu}_{\nu}(g_k)=0$. Through these steps, we derive the following constraints
$$
\nabla_{\mu}T^{\mu}_{\nu}(g_1)+P_0(1)\nabla_{\mu}\big[(S(1)-S(2))l^{\mu}l_{\nu} \big]=0,
$$
$$
 \nabla_{\mu}T^{\mu}_{\nu}(g_k)
        -P_0(k)\nabla_{\mu} \big[(S(k)-S(k+1))l^{\mu}l_{\nu}\big]+\frac{C^4(k-1)}{C^4(k)}P_0(k-1)\big[S(k-1)-S(k) \big]=0,
$$
\begin{equation}\label{gks2}        
         \nabla_{\mu}T^{\mu}_{\nu}(g_N)+P_0(N-1)\big[S(N-1)-S(N) \big]=0.
\end{equation}
As anticipated, there is no obligatory stipulation for the local conservation of stress-energy tensors. Nevertheless, we confine our analysis to scenarios in which such conservation is maintained. In this scenario, the only way to satisfy equation \eqref{gks2} is to implement the condition 
\begin{equation}\label{ksws1}
    P_0(k)=0,
\end{equation}
for all $j \in[1,N-1]$. Upon integrating this constraint into the equations of motion \eqref{gks2}, a simplification of the equations is achieved, as delineated below
$$
G^{\mu}_{\nu}(g_1)\kappa_1 T^{\mu}_{\nu}(g_1)+\frac{\kappa_1 m^2}{\kappa}\big[P_1(1)\delta^{\mu}_{\nu} \big], 
$$  
$$  
G^{\mu}_{\nu}(g_k)=\kappa_k T^{\mu}_{\nu}(g_k)+\frac{\kappa_k m^2}{\kappa}\big[P_1(k)\delta^{\mu}_{\nu}\big] +\frac{\kappa_k m^2}{\kappa}\frac{C^4(k-1)}{C^4(k)}\big[P_2(k-1)\delta^{\mu}_{\nu}\big], 
$$
\begin{equation}\label{kss2}   
    G^{\mu}_{\nu}(g_N)=\kappa_N T^{\mu}_{\nu}(g_N)+\frac{\kappa_N m^2}{\kappa}\frac{C^4(N-1)}{C^4(N)}\left[P_2(N-1)\delta^{\mu}_{\nu} \right].
\end{equation}
By incorporating the cosmological constants denoted as $\Lambda_k$ from equation \eqref{gks6} into equation \eqref{kss2} and performing a contraction with the metric tensor $(g_k)_{\mu \alpha}$, the equations of motion can be rewritten in compact form
\begin{equation}\label{kss3}
    G_{\mu \nu}(g_k)+\Lambda_k(g_k)_{ \mu \nu}=\kappa_kT_{\mu \nu}(g_k),
\end{equation}
for all $k \in[1,N]$.
Consequently, any Kerr-Schild ansatz that ensures the local conservation of the stress-energy tensor, satisfying Einstein's field equations, can be employed to construct a multigravity solution by using proportional spacetimes as follows:
\begin{equation}\label{ksws2}
    (g_j)_{\mu \nu} = C^2(j)\bigg[\overline{g}_{\mu \nu}+2S_j k_{\mu}k_{\nu} \bigg],
\end{equation}
for all $j \in[1,N]$ and 
$$
G_{\mu \nu}(g_j)+\Lambda_k(g_j)_{ \mu \nu}=\kappa_jT_{\mu \nu}(g_k), \ \ \  \forall \ k \in[1,N],
$$
$$
P_0(j)= 0, \ \ \ \forall \  j \in [1,N],
$$
\begin{equation}\label{ksws3}
\begin{split}
     \Lambda=&-\frac{\kappa_1m^2}{\kappa}P_1(1),\\
    \Lambda=&-\frac{\kappa_k m^2}{\kappa}\left(P_1(j)+\frac{C^4(k-1)}{C^4(j)}P_2(j-1) \right),\quad \forall \ j \in [2,N-1],\\
    \Lambda =&-\frac{\kappa_Nm^2}{\kappa}\frac{C^4(N-1)}{C^4(N)}P_2(N-1), 
\end{split}
\end{equation}
$$ 
    \nabla_{\mu}T^{\mu}_{\nu}(g_j)=0, 
$$ 
for all  $j \in[1,N]$. In the following subsections, we will construct solutions within the framework of multigravity that incorporate sources, adhering to the methodology outlined previously.
\subsubsection{Multi-Reissner-Nordström}
We begin the sourced solutions with the Reissner--Nordström metric, which solves the Einstein--Maxwell equations. In Kerr-Schild form (see \cite{Bah:2019sda,Ett:2015fhw}) it is written as
\begin{equation}\label{rs1}
    g_{\mu \nu}=\eta_{\mu \nu}+\left(\frac{2Gm}{r}-\frac{GQ^2}{4\pi\epsilon_0r^2}\right)k_{\mu}k_{\nu},
\end{equation}
with $k_{\mu}=(1,\widehat{r})$. Using this metric as the seed and applying the proportional Kerr-Schild prescription, we construct the corresponding multigravity configuration as follows
\begin{equation}\label{rs2}
    \begin{split}
        g_{\mu \nu}(j)=&C^{2}(j)\left[\eta_{\mu \nu}+\left(\frac{2G_jm_j}{r}-\frac{G_jQ_j^2}{4\pi\epsilon_{0j}r^2}\right)k_{\mu}k_{\nu} \right],\\
        P_0(j)=&0,\quad  \forall \ j \in [1,N],\\
    P_1(1) =&0, \\
    P_1(j)+\frac{C^4(j-1)}{C^4(j)}P_2(j)=&0 ,\quad \forall \ j \in [2,N-1],\\
    P_2(N-1)=&0.   
    \end{split}
\end{equation}
Equation \eqref{rs2} describes a multigravity solution consisting of $N$ Reissner--Nordström black holes, with independent masses $m_j$ and electric charges $Q_j$ in each sector. In addition, this family can be recovered as an appropriate limiting case of the Kerr-Newman-AdS metric in \cite{Wood:2024acv}.
\subsubsection{Multi-Reissner-Nordström-AdS}
To incorporate a cosmological constant, we may instead start from the Reissner--Nordström--Anti-de Sitter (AdS) metric. As explained in \cite{Hawking:1973uf}, its Kerr-Schild representation is
\begin{equation}\label{rsads1}
    g_{\mu \nu}=\bar{g}_{\mu \nu}+\left(\frac{2Gm}{r}-\frac{GQ^2}{4\pi\epsilon_0r^2}\right)k_{\mu}k_{\nu},
\end{equation}
where the AdS background is
\begin{equation}\label{rsads2}
    \overline{g}_{\mu \nu} ={\rm diag} \left(1-\frac{\Lambda r^2}{3}, \left(1-\frac{\Lambda r^2}{3}\right)^{-1},r^2,r^2 \sin^2\theta\right)
\end{equation}
and the null vector is
\begin{equation}\label{rsads3}
    k_{\mu}=\left(1,\frac{1}{1-\frac{\Lambda r^2}{3}},0,0\right).
\end{equation}
Using this AdS seed, we can now write down the associated multigravity solution as
$$
g_{\mu \nu}(j)=C^2(j)\left[\overline{g}_{\mu \nu}+\left(\frac{2G_jm_j}{r}-\frac{G_jQ_j^2}{4\pi\epsilon_0r^2}\right)k_{\mu}k_{\nu}\right],
$$
$$
P_0(j)=0,\quad \forall \ j \in [1,N],
$$
\begin{equation}\label{rsads4}
    \begin{split}
     \Lambda=&-\frac{\kappa_1m^2}{\kappa}P_1(1),\\
    \Lambda=&-\frac{\kappa_k m^2}{\kappa}\left(P_1(j)+\frac{C^4(k-1)}{C^4(j)}P_2(j-1) \right),\quad \forall \ j \in [2,N-1],\\
    \Lambda =&-\frac{\kappa_Nm^2}{\kappa}\frac{C^4(N-1)}{C^4(N)}P_2(N-1).   
    \end{split}
\end{equation}
Then this multigravity solution \eqref{rsads4} is formed by \(N\) Reissner--Nordström Anti-de Sitter (AdS) black holes. The masses \(m_j\) and electric charges \(Q_j\) may differ from sector to sector, while all metrics are supported by the same AdS background and hence share a common cosmological constant.
\subsubsection{Multi-Kerr-Newman}
As a rotating and charged generalization, consider the Kerr--Newman metric in Kerr-Schild form (see \cite{Newman:1965my,Debney:1969zz})
\begin{equation}\label{kn1}
    g_{\mu \nu}=\eta_{\mu \nu}+\frac{1}{\Sigma}\left(2Gmr-\frac{GQ^2}{4\pi\epsilon_0} \right)k_{\mu}k_{\nu},
\end{equation}
where
\begin{equation}\label{kn2}
    \Sigma=r^2+a^2\cos^2\theta,\ \ \  k_{\mu}=\left(1,\frac{rx+ay}{r^2+a^2},\frac{ry-ax}{r^2+a^2},\frac{z}{r} \right).
\end{equation}
Taking \eqref{kn1}--\eqref{kn2} as the seed metric, the proportional Kerr-Schild construction leads to the following multigravity solution
\begin{equation}\label{knf}
    \begin{split}
         g_{\mu \nu}(j)=&C^2(j)\left[\eta_{\mu \nu}+\frac{1}{\Sigma}\left(2G_jm_jr-\frac{G_jQ_j^2}{4\pi\epsilon_{0j}} \right)k_{\mu}k_{\nu}\right],\\
                 P_0(j)=&0,\quad \forall \ j \in [1,N],\\
    P_1(1) =&0, \\
    P_1(j)+\frac{C^4(j-1)}{C^4(j)}P_2(j)=&0, \quad \forall \  j \in[2,N-1],\\
    P_2(N-1)=&0. 
    \end{split}
\end{equation}
Equation \eqref{knf} describes a multigravity solution comprising $N$ Kerr--Newman black holes. Although the angular momentum parameter is common in all sectors, the masses $m_j$ and the electric charges $Q_j$ are allowed to be distinct. This family is obtained as a limiting case of the Kerr-Newman-AdS metric presented in \cite{Wood:2024acv} (derived there using the vielbein formalism) and is rederived here in the metric formulation. 
\subsubsection{Multi-Kerr-Newman-AdS}
Finally, we include both rotation, charge, and a cosmological constant by considering the Kerr--Newman--AdS metric (see \cite{Gibbons:2004uw,Malek:2010mh}),
\begin{equation}\label{knads1}
    g_{\mu \nu}=\overline{g}_{\mu \nu}+\frac{2Gmr-\frac{GQ^2}{4\pi \epsilon_0}}{\Sigma},
\end{equation}
where
$$
    \Sigma=r^2+a^2\cos^2\theta, \ \ \ \ \  \Delta_{\theta}=1+\frac{\Lambda}{3}a^2\cos^2\theta,
$$
\begin{equation}\label{knads2}
    \Xi = 1+\frac{\Lambda}{3}a^2,\ \ \ \ \
    \Delta_r=(r^2+a^2)\left(1-\frac{\Lambda}{3}r^2\right)
\end{equation}
and
\begin{equation}\label{knads3}
    k_{\mu}= \left(1,\frac{\Sigma}{\Delta_r},0,-\frac{a^2\sin^2\theta}{\Xi} \right).
\end{equation}
With this Kerr--Newman-AdS seed at hand, the multigravity extension follows from the same proportional prescription and is written as
$$
g_{\mu \nu}(j)=C^2(j)\left[\overline{g}_{\mu \nu}+\frac{2G_jm_jr-\frac{G_jQ_j^2}{4\pi \epsilon_0}}{\Sigma}\right],
$$
$$
P_0(j)=0, \ \ \ \  \forall \ j \in [1,N],
$$
\begin{equation}\label{knads4}
\begin{split}
     \Lambda=&-\frac{\kappa_1m^2}{\kappa}P_1(1),\\
    \Lambda=&-\frac{\kappa_k m^2}{\kappa}\left(P_1(j)+\frac{C^4(k-1)}{C^4(j)}P_2(j-1) \right),\quad \forall \ j \in [2,N-1],\\
    \Lambda =&-\frac{\kappa_Nm^2}{\kappa}\frac{C^4(N-1)}{C^4(N)}P_2(N-1).
\end{split}
\end{equation}
This previous equation \eqref{knads4} describes a multigravity solution made of $N$ Kerr--Newman--AdS black holes. The angular momentum parameter is common to all sectors, whereas the masses $m_j$ and electric charges $Q_j$ remain independent, and all metrics live on the same AdS background. This solution was first derived in Ref. \cite{Wood:2024acv} using the vielbein formalism and is also rederived here using the metric formulation.

\vskip 1truecm
\section{Proportional Double Kerr-Schild in Multigravity}\label{S4}
In this section, we consider solutions in multigravity by extending our ansatz to a double Kerr-Schild ansatz. In this context, the double Kerr-Schild metric takes the form
\begin{equation}\label{dks1}
    g_{\mu \nu} = \overline{g}_{\mu \nu}+2Sl_{\mu}l_{\nu}+2Qf_{\mu}f_{\nu},
\end{equation}
where $l^2=f^2=l\cdot f=0$. The inverse metric takes the following form
\begin{equation}\label{dks2}
    g^{\mu \nu}=\overline{g}^{\mu \nu}-2Sl^{\mu}l^{\nu}-2Qf^{\mu}f^{\nu}.
\end{equation}
This particular ansatz can be employed to formulate a proportional double Kerr-Schild ansatz within multigravity as follows
\begin{equation}\label{dks3}
    (g_j)_{\mu \nu}=C(j)^2\bigg[\overline{g}_{\mu \nu}+2S(j)l_{\mu}l_{\nu}+2Q(j)f_{\mu}f_{\nu}\bigg],
\end{equation}
where $l^2=0$ and $C(1)=1$.
A key property of the Kerr-Schild ansatz is its linearity in the inverse metrics
\begin{equation}\label{dks4}
 (g_j)^{\mu \nu}=\frac{1}{C(j)^2}\bigg[\overline{g}^{\mu \nu}-2S(j)l^{\mu}l^{\nu}-2Q(j)f^{\mu}f^{\nu} \bigg].
\end{equation}
In this case, the structure of the gamma matrix is still linear
\begin{equation}\label{dks5}
    (\gamma(g_j,g_{j+1})^n)^{\mu}_{\nu}=\frac{C(j+1)^{n}}{C(j)^n}\bigg[\delta^{\mu}_{\nu}-n(S(j)-S(j+1))l^{\mu}l_{\nu}-n(Q(j)-Q(j+1))f^{\mu}f_{\nu} \bigg].
\end{equation}
Consequently, the equations of motion for the double Kerr-Schild ansatz are expressed as follows
\begin{equation}\label{dks6}
    \begin{split}
    G^{\mu}_{\nu}(g_1)=&\kappa_1 T^{\mu}_{\nu}(g_1)+\frac{\kappa_1 m^2}{\kappa}\bigg[P_1(1)\delta^{\mu}_{\nu}-P_0(1)\big(S(1)-S(2)\big)l^{\mu}l_{\nu}-P_0(1)\big(Q(1)-Q(2)\big)f^{\mu}f_{\nu} \bigg], \\
    G^{\mu}_{\nu}(g_k)=&\kappa_k T^{\mu}_{\nu}(g_k)+\frac{\kappa_k m^2}{\kappa}\bigg[P_1(k)\delta^{\mu}_{\nu}-P_0(k)\big(S(k)-S(k+1)\big)l^{\mu}l_{\nu}-P_0(k)\big(Q(k)-Q(k+1)\big)f^{\mu}f_{\nu}\bigg] \\&+\frac{\kappa_k m^2}{\kappa}\frac{C^4(k-1)}{C^4(k)}\bigg[P_2(k-1)\delta^{\mu}_{\nu}+P_0(k-1)\big(S(k-1)-S(k)\big)l^{\mu}l_{\nu} \\&+P_0(k-1)\big(Q(k-1)-Q(k)\big)f^{\mu}f_{\nu}\bigg], \\
    G^{\mu}_{\nu}(g_N)=&\kappa_N T^{\mu}_{\nu}(g_N)+\frac{\kappa_N m^2}{\kappa}\frac{C^4(N-1)}{C^4(N)}\bigg[P_2(N-1)\delta^{\mu}_{\nu}+P_0(N-1)\big(S(N-1)-S(N)\big)l^{\mu}l_{\nu}\\&+P_0(N-1)\big(Q(N-1)-Q(N)\big) \bigg].
    \end{split}
\end{equation}
Focusing again on the scenario of local conservation of energy-momentum tensors, where $\nabla_{\mu}T^{\mu \nu}(g_k)=0$, equation \eqref{dks6} reduces to
$$
G^{\mu}_{\nu}(g_1)=\kappa_1 T^{\mu}_{\nu}(g_1)+\frac{\kappa_1 m^2}{\kappa}\big[P_1(1)\delta^{\mu}_{\nu} \big],
$$
$$
G^{\mu}_{\nu}(g_k)=\kappa_k T^{\mu}_{\nu}(g_k)+\frac{\kappa_k m^2}{\kappa}\big[P_1(k)\delta^{\mu}_{\nu}\big] +\frac{\kappa_k m^2}{\kappa}\frac{C^4(k-1)}{C^4(k)}\big[P_2(k-1)\delta^{\mu}_{\nu}\big],
$$
\begin{equation}\label{dks7}
    G^{\mu}_{\nu}(g_N)=\kappa_N T^{\mu}_{\nu}(g_N)+\frac{\kappa_N m^2}{\kappa}\frac{C^4(N-1)}{C^4(N)}\big[P_2(N-1)\delta^{\mu}_{\nu} \big].
\end{equation}
By contracting equation \eqref{dks7} with the tensor $(g_k)_{\mu \alpha}$, we can rewrite equations of motion as follows
\begin{equation}\label{dks8}
    G_{\mu \nu}(g_k)+\Lambda_kg_{k \mu \nu}=\kappa_k T_{\mu \nu}(g_k).
\end{equation}
Consequently, any double Kerr-Schild ansatz that ensures the local conservation of the stress-energy tensor, satisfying Einstein equations, can be employed to construct a multigravity solution by using proportional spacetimes is given by
\begin{equation}\label{dks9}
    \begin{split}
    (g_j)_{\mu \nu} = C^2(j)\bigg[\overline{g}_{\mu \nu}+2S_j k_{\mu}k_{\nu}+2Q_j l_{\mu}l_{\nu} \bigg], 
    \end{split}
\end{equation}
for all $j \in [1,N]$ and where
$$
G_{\mu \nu}(g_j)+\Lambda_k(g_j)_{ \mu \nu}=\kappa_jT_{\mu \nu}(g_k), \ \ \  \forall \ k \in[1,N],
$$
$$
P_0(j)=0, \ \ \  \forall \ j \in [1,N],
$$
\begin{equation}\label{dks10}
\begin{split}
     \Lambda=&-\frac{\kappa_1m^2}{\kappa}P_1(1),\\
    \Lambda=&-\frac{\kappa_k m^2}{\kappa}\left(P_1(j)+\frac{C^4(k-1)}{C^4(j)}P_2(j-1) \right),\quad \forall \  j \in [2,N-1],\\
    \Lambda =&-\frac{\kappa_Nm^2}{\kappa}\frac{C^4(N-1)}{C^4(N)}P_2(N-1),  
\end{split}
\end{equation}
$$
 \nabla_{\mu}T^{\mu}_{\nu}(g_j)=0, \ \ \  \forall \ j \in[1,N]. 
$$
In the following subsections, we will construct solutions within the framework of multigravity adhering to the methodology outlined previously.

\subsection{Multi-Taub-NUT}
The Taub-NUT-Kerr-deSitter metric is expressed \cite{Taub:1950ez,Newman:1963yy,Plebanski:1975xfb} in the double Kerr-Schild form by
\begin{equation}\label{tn1}
    g_{\mu \nu}=\overline{g}_{\mu \nu}+\kappa\left[\frac{2Np}{q^2-p^2}k_{\mu}k_{\nu}+\frac{2Mq}{q^2-p^2}l_{\mu}l_{\nu} \right].
\end{equation}
The background metric is given by the line element
\begin{equation}
\overline{ds}^{2}
= -\frac{1}{q^{2}-p^{2}}
\Big[
\overline{\Delta}_{p}\,(d\widetilde{\tau} + q^{2} d\widetilde{\sigma})^{2}
- \overline{\Delta}_{q}\,(d\widetilde{\tau} + p^{2} d\widetilde{\sigma})^{2}
\Big]
- 2\,(d\widetilde{\tau} + q^{2} d\widetilde{\sigma})\,dp
- 2\,(d\widetilde{\tau} + p^{2} d\widetilde{\sigma})\,dq,
\end{equation}
where
\begin{equation}
\overline{\Delta}_{p} = \gamma - \epsilon p^{2} + \lambda p^{4},
\qquad
\overline{\Delta}_{q} = -\gamma + \epsilon q^{2} - \lambda q^{4}.
\end{equation}
In the coordinate system denoted by $(\widetilde{\tau}, \widetilde{\sigma}, p, q)$, which has a $(2,2)$ signature, the Kerr-Schild null vectors can be expressed as follows
\begin{equation}
k_{\mu} = (1,\, q^{2},\, 0,\, 0), \qquad
l_{\mu} = (1,\, p^{2},\, 0,\, 0).
\end{equation}
Subsequently, we are able to construct the corresponding multigravity solution in the following manner
$$
g_{\mu \nu}(j)=C^2(j)\left[\overline{g}_{\mu \nu}+\kappa_j\left(\frac{2N_jp}{q^2-p^2}k_{\mu}k_{\nu}+\frac{2M_jq}{q^2-p^2}l_{\mu}l_{\nu} \right)\right],
$$
$$
P_0(j)= 0,\quad \forall \ j \in [1,N],
$$
\begin{equation}\label{tnads1}
    \begin{split}
     \Lambda=&-\frac{\kappa_1m^2}{\kappa}P_1(1),\\
    \Lambda=&-\frac{\kappa_k m^2}{\kappa}\left(P_1(j)+\frac{C^4(k-1)}{C^4(j)}P_2(j-1) \right),\quad \forall \ j \in [2,N-1],\\
    \Lambda =&-\frac{\kappa_Nm^2}{\kappa}\frac{C^4(N-1)}{C^4(N)}P_2(N-1).
    \end{split}
\end{equation}
Equation \eqref{tnads1} describes a solution in multigravity, representing a collection of $N$ Taub-Nut black holes, each characterized by independent $M,N$ parameters.

\subsection{Multi-Plebański-Demiański}
The Plebański-Demiański metric in double Kerr-Schild form  is given by \cite{Plebanski:1975xfb,Plebanski:1976gy}
\begin{equation}\label{pd1}
g_{\mu \nu}=\overline{g}_{\mu \nu}+\frac{\kappa}{2}\left[\frac{2Np+G^2}{8\pi(q^2-p^2)}k_{\mu}k_{\nu}+\frac{2Mq-Q^2}{8\pi(q^2-p^2)}l_{\mu}l_{\nu} \right].
\end{equation}
The background metric is given by the line element
\begin{equation}\label{pd2}
\overline{ds}^{2}
= -\frac{1}{q^{2}-p^{2}}
\Big[
\overline{\Delta}_{p}\,(d\widetilde{\tau} + q^{2} d\widetilde{\sigma})^{2}
- \overline{\Delta}_{q}\,(d\widetilde{\tau} + p^{2} d\widetilde{\sigma})^{2}
\Big]
- 2\,(d\widetilde{\tau} + q^{2} d\widetilde{\sigma})\,dp
- 2\,(d\widetilde{\tau} + p^{2} d\widetilde{\sigma})\,dq,
\end{equation}
where
\begin{equation}\label{pd3}
\overline{\Delta}_{p} = \gamma - \epsilon p^{2} + \lambda p^{4},
\qquad
\overline{\Delta}_{q} = -\gamma + \epsilon q^{2} - \lambda q^{4}.
\end{equation}

The Kerr-Schild null vectors are given in the
$(\widetilde{\tau}, \widetilde{\sigma}, p, q)$ coordinate system
(with $(2,2)$ signature) by
\begin{equation}\label{pd4}
k_{\mu} = (1,\, q^{2},\, 0,\, 0), \qquad
l_{\mu} = (1,\, p^{2},\, 0,\, 0).
\end{equation}
Thus, we can construct the corresponding multigravity solution which is given by
$$
g_{\mu \nu}(j)=C^2(j)\left[\overline{g}_{\mu \nu}+\frac{\kappa_j}{2}\left(\frac{2N_jp+G_j^2}{8\pi(q^2-p^2)}k_{\mu}k_{\nu}+\frac{2M_jq-Q_j^2}{8\pi(q^2-p^2)}l_{\mu}l_{\nu} \right)\right],
$$
$$
P_0(j)=0, \ \ \ \forall \ j \in [1,N],
$$
\begin{equation}\label{pds1}
    \begin{split}    
     \Lambda=&-\frac{\kappa_1m^2}{\kappa}P_1(1),\\
    \Lambda=&-\frac{\kappa_k m^2}{\kappa}\left(P_1(j)+\frac{C^4(k-1)}{C^4(j)}P_2(j-1) \right), \ \ \  \forall \ j \in [2,N-1],\\
    \Lambda =&-\frac{\kappa_Nm^2}{\kappa}\frac{C^4(N-1)}{C^4(N)}P_2(N-1).
    \end{split}
\end{equation}
This solution \eqref{pds1} describes a collection of $N$ Plebański-Demiański metrics, each with distinct parameters $M,N,G$.

\vskip 1truecm
\section{Classical Double Copy in Multigravity}\label{S5}
As was mentioned in the introduction section, the {\it classical double copy} establishes a correspondence between gauge and gravity solutions of the corresponding classical field equations. Thus the classical double copy connects solutions from classical gauge theory to those in GR, as discussed in \cite{Monteiro:2014cda} (and reviewed in \cite{White:2024pve}). This methodology has been broadened to include the bigravity scenario \cite{Garcia-Compean:2024zze,Garcia-Compean:2024uie}. While it is not constrained to this, we will focus on studying and expanding it within the multigravity context, specifically for the proportional Kerr-Schild solutions described in earlier sections. Let's start with a Kerr-Schild metric defined by
\begin{equation}\label{dc1}
    g_{\mu \nu}=\overline{g}_{\mu \nu}+\phi k_{\mu}k_{\nu},
\end{equation}
where $k\cdot k=0$. According to \cite{Bah:2019sda,Carrillo-Gonzalez:2017iyj}, the mixed index Ricci tensor for this spacetime can be expressed as follows
\begin{equation}\label{dc2}
    R^{\alpha}_{\beta}(g)=\overline{R}^{\alpha}_{\beta}(\overline{g})-\phi k^{\alpha}k^{\sigma}\overline{R}_{\sigma \beta}(\overline{g})+\frac{1}{2}\overline{\nabla}_{\sigma}\bigg[\overline{\nabla}^{\alpha}(\phi k^{\sigma}k_{\beta})+\overline{\nabla}_{\beta}(\phi k^{\sigma}k^{\alpha})-\overline{\nabla}^{\sigma}(\phi k^{\alpha}k_{\beta}) \bigg].
\end{equation}
The spin-1 gauge field $A^{\mu}$ and its corresponding strength tensor $F^{\mu \nu}$ are defined as follows
\begin{equation}\label{dc3}
\begin{split}
    A^{\mu}=&\phi k^{\mu},\\
    F^{\alpha \beta}=&\overline{\nabla}^{\alpha}A^{\beta}-\overline{\nabla}^{\beta}A^{\alpha}.
\end{split}
\end{equation}
We can then reformulate Eq. \eqref{dc2} as
\begin{equation}\label{dc4}
    R^{\mu}_{\nu}=\overline{R}^{\mu}_{\nu}-\frac{1}{2}\left[\overline{\nabla}_{\lambda}F^{\lambda \mu}+\frac{\overline{R}}{6}A^{\mu} \right]k_{\nu}-\frac{1}{2}\left[X^{\mu}_{\nu}+Y^{\mu}_{\nu}\right],
\end{equation}
where
\begin{equation}\label{dc5}
    \begin{split}
    X^{\mu}_{\nu}=& -\overline{\nabla}_{\nu}\left[A^{\mu}\left(\overline{\nabla}_{\lambda}k^{\lambda}+\frac{k^{\lambda}\overline{\nabla}_{\lambda}\phi}{\phi}\right) \right],\\
    Y^{\mu}_{\nu}=&F^{\rho \mu}\overline{\nabla}_{\rho}k_{\nu}-\overline{\nabla}_{\rho}\left(A^{\rho}\overline{\nabla}^{\mu}k_{\nu}-A^{\mu}\overline{\nabla}^{\rho}k_{\nu} \right).
    \end{split}
\end{equation}
The multigravity equations of motion for proportional Kerr-Schild spacetimes can be reformulated in a trace-reversed manner as follows
$$
R^{\mu}_{\nu}(g_1)=\kappa_1 \bigg[T^{\mu}_{\nu}(g_1)-\frac{1}{2}T(g_1)\delta^{\mu}_{\nu}\bigg]-\frac{\kappa_1 m^2}{\kappa}\big[P_1(1)\delta^{\mu}_{\nu} \big], 
$$
$$
R^{\mu}_{\nu}(g_k)=\kappa_k \left[T^{\mu}_{\nu}(g_k)-\frac{1}{2}T(g_k)\delta^{\mu}_{\nu}\right]-\frac{\kappa_k m^2}{\kappa}\big[P_1(k)\delta^{\mu}_{\nu}\big]-\frac{\kappa_k m^2}{\kappa}\frac{C^4(k-1)}{C^4(k)}\big[P_2(k-1)\delta^{\mu}_{\nu}\big],
$$
\begin{equation}\label{dc6}
    R^{\mu}_{\nu}(g_N)=\kappa_N\left[ T^{\mu}_{\nu}(g_N)-\frac{1}{2}T(g_N)\delta^{\mu}_{\nu}\right]-\frac{\kappa_N m^2}{\kappa}\frac{C^4(N-1)}{C^4(N)}\big[P_2(N-1)\delta^{\mu}_{\nu} \big].
\end{equation}
Substituting the derived expression for the Ricci tensor into Eq. \eqref{dc2}, we derive the subsequent equations of motion
$$
\overline{R}^{\mu}_{\nu}(g_1)-\frac{1}{2}\left[\overline{\nabla}_{\lambda}F^{\lambda \mu}(A_1)+\frac{\overline{R}(g_1)}{6}A^{\mu}_1 \right]k_{\nu}-\frac{1}{2}\left[X^{\mu}_{\nu}(A_1)+Y^{\mu}_{\nu}(A_1)\right]
$$
$$
=\kappa_1 \left[T^{\mu}_{\nu}(g_1)-\frac{1}{2}T(g_1)\delta^{\mu}_{\nu}\right] -\frac{\kappa_1 m^2}{\kappa}\left[P_1(1)\delta^{\mu}_{\nu} \right], 
$$
$$
\overline{R}^{\mu}_{\nu}(g_k)-\frac{1}{2}\left[\overline{\nabla}_{\lambda}F^{\lambda \mu}(A_k)+\frac{\overline{R}(g_k)}{6}A^{\mu}_k \right]k_{\nu}-\frac{1}{2}\left[X^{\mu}_{\nu}(A_k)+Y^{\mu}_{\nu}(A_k)\right]
$$
$$
=\kappa_k \left[T^{\mu}_{\nu}(g_k)-\frac{1}{2}T(g_k)\delta^{\mu}_{\nu}\right]-\frac{\kappa_k m^2}{\kappa}\left[P_1(k)\delta^{\mu}_{\nu}\right]-\frac{\kappa_k m^2}{\kappa}\frac{C^4(k-1)}{C^4(k)}\left[P_2(k-1)\delta^{\mu}_{\nu}\right],
$$
$$
\overline{R}^{\mu}_{\nu}(g_N)-\frac{1}{2}\left[\overline{\nabla}_{\lambda}F^{\lambda \mu}(A_N)+\frac{\overline{R}(g_N)}{6}A^{\mu}_N \right]k_{\nu}-\frac{1}{2}\left[X^{\mu}_{\nu}(A_N)
    +Y^{\mu}_{\nu}\right](A_N)
$$
\begin{equation}\label{dc7}     
=\kappa_N\left[ T^{\mu}_{\nu}(g_N)-\frac{1}{2}T(g_N)\delta^{\mu}_{\nu}\right]-\frac{\kappa_N m^2}{\kappa}\frac{C^4(N-1)}{C^4(N)}\left[P_2(N-1)\delta^{\mu}_{\nu} \right].
\end{equation}
The proportional Kerr-Schild solutions obtained fulfil the $\Lambda_k$ constraints
\begin{equation}\label{dc8}
\begin{split}
         \Lambda_1=&-\frac{\kappa_1m^2}{\kappa}P_1(1),\\
    \Lambda_k=&-\frac{\kappa_k m^2}{\kappa}\left(P_1(k)+\frac{C^4(k-1)}{C^4(k)}P_2(k-1) \right),\\
    \Lambda_N =&-\frac{\kappa_Nm^2}{\kappa}\frac{C^4(N-1)}{C^4(N)}P_2(N-1). 
\end{split}
\end{equation}
In addition the background metrics $\overline{g}_{\mu \nu}(k)$ satisfies
\begin{equation}\label{dc9}
    \overline{R}^{\mu}_{\nu}(\overline{g}_k)=\Lambda_k \delta^{\mu}_{\nu}.
\end{equation}
Thus the equations of motion \eqref{dc7} can be expressed in the form \begin{equation}\label{dc10}
    \begin{split}
    -\frac{1}{2}\left[\overline{\nabla}_{\lambda}F^{\lambda \mu}(A_1)+\frac{\overline{R}(g_1)}{6}A^{\mu}_1 \right]k_{\nu}-\frac{1}{2}\big[X^{\mu}_{\nu}(A_1)+Y^{\mu}_{\nu}(A_1)\big]=&\kappa_1 \left[T^{\mu}_{\nu}(g_1)-\frac{1}{2}T(g_1)\delta^{\mu}_{\nu}\right], \\
    -\frac{1}{2}\left[\overline{\nabla}_{\lambda}F^{\lambda \mu}(A_k)+\frac{\overline{R}(g_k)}{6}A^{\mu}_k \right]k_{\nu}-\frac{1}{2}\big[X^{\mu}_{\nu}(A_k)+Y^{\mu}_{\nu}(A_k)\big]=&\kappa_k \left[T^{\mu}_{\nu}(g_k)-\frac{1}{2}T(g_k)\delta^{\mu}_{\nu}\right], \\
    -\frac{1}{2}\left[\overline{\nabla}_{\lambda}F^{\lambda \mu}(A_N)+\frac{\overline{R}(g_N)}{6}A^{\mu}_N \right]k_{\nu}-\frac{1}{2}\big[X^{\mu}_{\nu}(A_N)+Y^{\mu}_{\nu}(A_N)\big]=&\kappa_N\left[ T^{\mu}_{\nu}(g_N)-\frac{1}{2}T(g_N)\delta^{\mu}_{\nu}\right].
    \end{split}
    \end{equation}
By contracting equations \eqref{dc10} with a Killing vector $V^{\mu}(k)$, which is a common Killing vector field for both metrics $g(k)$ and $\overline{g}(k)$, we derive the equation of motion pertinent to the single copy fields denoted as $A_k^{\mu}$:
\begin{equation}\label{dc13}
    \begin{split}
    \overline{\nabla}_{\lambda}F^{\lambda \mu}(A_k)+\frac{\overline{R}(g_k)}{6}A^{\mu}_k +\frac{V^{\nu}}{V^{\alpha}k_{\alpha}}\left[X^{\mu}_{\nu}(A_k)+Y^{\mu}_{\nu}(A_k)\right]=  J^{\mu}_k,
    \end{split}
    \end{equation}
where
\begin{equation}\label{dc12}
    J^{\mu}_k=-\frac{2V^{\nu}}{V^{\alpha}k_{\alpha}}\kappa_k \left[T^{\mu}_{\nu}(g_k)-\frac{1}{2}T(g_k)\delta^{\mu}_{\nu}\right].
\end{equation}
Equation \eqref{dc13} is the equation of motion satisfied by the single-copy spin-1 field $A^{\mu}_{k}$.\\
By contracting again with $V_{\mu}(k)$, we derive the equations of motion for the zero-copy fields
\begin{equation}\label{dc14}
\overline{\nabla}^2\phi_k+\frac{\overline{R}(g_k)}{6}\phi_k=j_k-\frac{V_{\nu}}{(V^{\alpha}_k k_{\alpha})^2}\left(V^{\mu}_kX^{\nu}_{\mu}(g_k) +V^{\mu}_kY^{\nu}_{\mu}(g_k)+Z^{\nu}\right),
\end{equation}
where
\begin{equation}\label{dc15}
    j_k=\frac{V_{\nu}(k)J^{\nu}_{k}}{V^{\rho}_k k_{\rho}}, \ \ \ \ \  Z^{\nu}_k=V^{\rho}_k k_{\rho}\overline{\nabla}_{\mu}\left(\phi_k\overline{\nabla}^{[\mu}k^{\nu]}-k^{\mu}\overline{\nabla}_{\nu}\phi_k \right).
\end{equation}
In this section, we will concentrate on developing the double copy for stationary proportional Kerr-Schild spacetimes. For these types of spacetimes, a Killing vector is given by
\begin{equation}\label{dc16}
    V^{\mu}(g_k)=1.
\end{equation}
Consequently, the equations of motion for the single and zero copies are simplified and it yields 
\begin{equation}\label{dc17}
    \begin{split}
    \overline{\nabla}_{\lambda}F^{\lambda \mu}(A_k)+\frac{\overline{R}(g_k)}{6}A^{\mu}_k =J^{\mu}_k,
    \end{split}    
\end{equation}
\begin{equation}\label{dc18}
\overline{\nabla}^2\phi_k+\frac{\overline{R}(g_k)}{6}\phi_k=j_k.
\end{equation}
Consequently, the single-copy fields satisfy a Proca equation of motion characterized by a mass term  $m^2 =- \frac{\overline{R}(g_k)}{6}$, which is induced by the background Ricci scalar. In this context, the single copy fields can be interpreted as massive photons, with their associated mass arising from the scalar curvature of the metrics. Massless photons are derived from spacetimes characterized by a null Ricci scalar. 

Similarly, the zero-copy fields are governed by a Klein-Gordon equation with an equivalent mass term \( m^2 =- \frac{\overline{R}(g_k)}{6} \), also determined by the background Ricci scalar. In this context, the zero copy fields can be interpreted as real massive scalar fields, with their mass associated to the scalar curvature of the metrics. Massless scalar fields are also derived from spacetimes characterized by a null Ricci scalar.
In the following subsections we will explore the single and zero copy fields that can be generated form the proportional Kerr-Schild solutions in multigravity found in previous sections.

In order to make contact with a standard field-theory description, we note that the equations of motion obeyed by the single-copy spin-1 fields $A_k$ can be obtained from an interacting Abelian theory of $N$ gauge fields on the background spacetime. A convenient starting point is the quadratic Lagrangian
\begin{equation}\label{Ab1}
    \mathcal{L}_{N} = -\frac{1}{4}\sum_{n=1}^{N}F^{(n)}_{\mu \nu}F^{(n)\mu \nu}-\frac{1}{2}\sum_{n,m}I_{nm}A_{\mu}^{(n)}A_{\nu}^{(m)}\bar{g}^{\mu \nu}.
\end{equation}
Equation \eqref{Ab1} describes an Abelian gauge theory with $N$ interacting spin-1 fields $A_{\mu}^{(n)}$ in curved spacetime. It is manifestly invariant under $N$ independent $U(1)$ gauge symmetries, so the symmetry group is $U(1)^N$. The interaction is quadratic in the fields and therefore produces linear (mass-like) terms in the equations of motion. We will work in a field basis in which the interaction matrix is diagonal,
\begin{equation}\label{Ab4}
    I_{nm}=M_n \delta_{nm}.
\end{equation}
To allow for external matter, we minimally couple each $A_{\mu}^{(n)}$ to a conserved source $J^{(n)\mu}$, so that
\begin{equation}\label{Ab11}
    \mathcal{L}_{N} = -\frac{1}{4}\sum_{n=1}^{N}F^{(n)}_{\mu \nu}F^{(n)\mu \nu}-\frac{1}{2}\sum_{n,m}I_{nm}A_{\mu}^{(n)}A_{\nu}^{(m)}\bar{g}^{\mu \nu}+\sum_{n=1}^{N}A_{\mu}^{(n)}J^{(n)\mu}.
\end{equation}
While a detailed discussion of phenomenology is beyond the scope of this work, such multi-vector field models appear, for instance, in the dark-sector constructions involving dark photons; see \cite{Fabbrichesi2021}. Varying the action yields the equations of motion
\begin{equation}\label{Ab3}
    \begin{split}
    \overline{\nabla}_{\lambda}F^{\lambda \mu}(A_k)-M_kA^{\mu}_k =J^{\mu}_k.
    \end{split}    
\end{equation}
Equation \eqref{Ab3} matches the single-copy equation \eqref{dc17} provided we restrict to equal Proca masses
\begin{equation}\label{Ab45}
    M_k=-\frac{\bar{R}(g_k)}{6}.
\end{equation}
Hence, the single-copy fields $A_k$ admit a natural interpretation as Proca fields (massive photons) emerging from a quadratic $U(1)^N$ theory. In this way, the single-copy map relates the multigravity Kerr--Schild sector to solutions of a $U(1)^N$ Proca gauge theory.

A completely analogous construction applies to the zero-copy scalars. The equations obeyed by $\phi_k$ in Eq.~\eqref{dc18} can be derived from a quadratic interacting multi-scalar theory with Lagrangian
\begin{equation}\label{scf1}
    \mathcal{L}_{\phi}=-\frac{1}{2}\sum_{n=1}^{N}\nabla_{\mu}\phi_n\nabla_{\nu}\phi_n\bar{g}^{\mu \nu}-\frac{1}{2}\sum_{n,m}V_{nm}\phi_{n}\phi_{m}.
\end{equation}
The Lagrangian in Eq.~\eqref{scf1} is manifestly invariant under $N$ copies of $SO(3)$, so the symmetry group is $SO(3)^N$. As before, we work in a basis where the quadratic interaction matrix is diagonal, i.e.,
\begin{equation}\label{scf2}
    V_{nm}=M_n \delta_{nm},
\end{equation}
and we minimally couple each scalar to an external source $j_n$
\begin{equation}\label{scf3}
    \mathcal{L}_{\phi,j}=-\frac{1}{2}\sum_{n=1}^{N}\nabla_{\mu}\phi_n\nabla_{\nu}\phi_n\bar{g}^{\mu \nu}-\frac{1}{2}\sum_{n=1}^{N}M_n\phi_n^{2}+\sum_{n=1}^{N}\phi_n j_n.
\end{equation}
Although we will not pursue applications here, multi-scalar Lagrangians of this type are widely used in cosmology, particularly in multifield inflationary scenarios (see \cite{PhysRevD.110.L041302,Wands2007}). Varying the action gives
\begin{equation}\label{scf4}
    \bar{\nabla}^2\phi_k-M_k\phi_k=j_k.
\end{equation}
Imposing again a common mass
\begin{equation}\label{scf5}
    M_k=-\frac{\bar{R}(g_k)}{6}.
\end{equation}
Eq.~\eqref{scf4} becomes equivalent to Eq.~\eqref{dc18}. Therefore, the zero-copy fields $\phi_k$ can be interpreted as scalar fields arising from a quadratic $SO(3)^N$ interacting theory. With these field-theory identifications in place, we now turn to explicit examples of the single- and zero-copy fields associated with the proportional Kerr--Schild solutions discussed above.

\subsection{Example: Multi-Schwarzschild}
We use the Multi-Schwarzschild multigravity solution obtained in Section \ref{S3}; see Eq.~\eqref{sch2}. Reading off the corresponding single- and zero-copy fields, we find
\begin{equation}\label{schdc2}
    A^{\mu}_k =C^2(k)\frac{2G_k m_k}{r}k^{\mu}, \ \ \ \ \ \phi_k=C^2(k)\frac{2G_k m_k}{r}.
\end{equation}
Substituting Eq.~\eqref{schdc2} into the general equations of motion \eqref{dc17} and \eqref{dc18} yields
\begin{equation}\label{exsch2}
    \begin{split}
    \overline{\nabla}_{\lambda}F^{\lambda \mu}(A_k) =0,
    \end{split}    
\end{equation}
\begin{equation}\label{exsch3}
\overline{\nabla}^2\phi_k=0.
\end{equation}
Accordingly, $A^{\mu}_{k}$ describes a massless spin-1 field that solves Maxwell's equations in vacuum. It takes the form of an electromagnetic four-potential source based on a point charge $Q_k$ given as:
\begin{equation}\label{exsch4}
    Q_k=2C^2(k)G_km_k.
\end{equation}
Thus, the metric parameters $m_k$ and $G_k$ map directly into the electric charge of the single-copy solution. Likewise, $\phi_k$ solves the stationary homogeneous wave equation (Laplace's equation) and therefore plays the role of a massless spin-0 potential sourced by the same point charge
\begin{equation}\label{exsch5}
    Q_k=2C^2(k)G_km_k.
\end{equation}
This shows that the zero-copy field inherits $m_k$ and $G_k$ in the same way.
\subsection{Example: Multi-Schwarzschild-AdS}
Next, we turn to the Multi-Schwarzschild--AdS multigravity solution of Section \ref{S3}; see Eq.~\eqref{sch6} (and the definitions in Eqs.~\eqref{sch4}--\eqref{sch5}). The single- and zero-copy fields follow as
\begin{equation}\label{schadsdc4}
    A^{\mu}_k =C^2(k)\frac{\kappa_k m_k}{8\pi r}k^{\mu},\quad \phi_k=C^2(k)\frac{\kappa_k m_k}{8\pi r}.
\end{equation}
Inserting Eq.~\eqref{schadsdc4} into Eqs.~\eqref{dc17} and \eqref{dc18}, we obtain
\begin{equation}\label{schadsdc5}
    \begin{split}
    \overline{\nabla}_{\lambda}F^{\lambda \mu}(A_k)+\frac{2\Lambda}{3}A^{\mu}_k =0,
    \end{split}    
\end{equation}
\begin{equation}\label{schadsdc6}
\overline{\nabla}^2\phi_k+\frac{2\Lambda}{3}\phi_k=0.
\end{equation}
Equation \eqref{schadsdc5} shows that $A^{\mu}_{k}$ satisfies the vacuum Proca equation with the mass parameter $M_k=\frac{2\Lambda}{3}$. We may therefore regard $A^{\mu}_{k}$ as a massive spin-1 potential sourced by a point charge
\begin{equation}\label{schadsdc61}
    Q_k=\frac{\kappa_k}{2} C^2(k)m_k.
\end{equation}
Therefore $m_k$ and $\kappa_k$ set the charge, while $\Lambda$ fixes the Proca mass. Similarly, $\phi_k$ obeys the homogeneous Klein--Gordon equation with the same mass $M_k=\frac{2\Lambda}{3}$ and can be seen as a massive spin-0 potential sourced by
\begin{equation}\label{schadsdc7}
    Q_k=\frac{\kappa_k}{2} C^2(k)m_k,
\end{equation}
again with $\Lambda$ providing the field mass.
\subsection{Example: Multi-Reissner–Nordström}
Consider now the Multi-Reissner--Nordstr\"om multigravity solution of Section \ref{S3}; see Eq.~\eqref{rs2} (with the Kerr--Schild vector given below Eq.~\eqref{rs1}). Extracting the copy fields from Eq.~\eqref{rs2}, we get
\begin{equation}\label{rsdc2}
    A^{\mu}_j =C^2(j)\left(\frac{2G_jm_j}{r}-\frac{G_jQ_j^2}{4\pi\epsilon_{0j}r^2}\right)k^{\mu},\quad \phi_k=C^2(j)\left(\frac{2G_jm_j}{r}-\frac{G_jQ_j^2}{4\pi\epsilon_{0j}r^2}\right).
\end{equation}
When we plug Eq.~\eqref{rsdc2} into \eqref{dc17} and \eqref{dc18}, the fields satisfy the equations with an external source
\begin{equation}\label{rsdc3}
    \begin{split}
    \overline{\nabla}_{\lambda}F^{\lambda \mu}(A_k) =-2\kappa_kT^{\mu}_0(g_k),
    \end{split}    
\end{equation}
\begin{equation}\label{rsdc4}
\overline{\nabla}^2\phi_k=-2\kappa_kT^{0}_{0}(g_k),
\end{equation}
with
\begin{equation}\label{rsdc5}
    T^{0}_{0}(g_k)=-\frac{Q_k^2}{8\pi \epsilon_0r^4}, \ \ \ \ \ T^{i}_{0}(g_k)=0.
\end{equation}
Thus $A^{\mu}_{k}$ solves Maxwell's equations in the presence of the source $T^{\mu}_{0}(g_k)$ and may be read as an electromagnetic four-potential produced by a point charge $q_k$ together with a non-uniform electrostatic source $\rho_k$
\begin{equation}\label{rsdc6}
    q_k=2C^2(k)G_km_k, \ \ \ \ \ \rho_k=\frac{Q_k^2}{4\pi r^4}.
\end{equation}
In this mapping, $m_k$ and $G_k$ generate the point charge, while the Einstein--Maxwell stress-energy component $T^{0}_{0}(g_k)$ induces the spatially varying source term. The scalar field $\phi_k$ follows the analogous inhomogeneous Laplace equation with the same source $T^{0}_{0}(g_k)$, so it plays the role of a massless spin-0 electrostatic potential sourced by $q_k$ and $\rho_k$ as above.
\subsection{Example: Multi-Reissner-Nordstrom-AdS}
We next examine the Multi-Reissner--Nordstr\"om--AdS multigravity solution from Section \ref{S3}; see Eq.~\eqref{rsads4} (with background and Kerr--Schild vector in Eqs.~\eqref{rsads2}--\eqref{rsads3}). The single and zero-th copy fields take the form
\begin{equation}\label{rsadsdc4}
    A^{\mu}_k =C^2(j)\left(\frac{2G_jm_j}{r}-\frac{G_jQ_j^2}{4\pi\epsilon_{0j}r^2}\right)k^{\mu},\quad \phi_k=C^2(j)\left(\frac{2G_jm_j}{r}-\frac{G_jQ_j^2}{4\pi\epsilon_{0j}r^2}\right).
\end{equation}
From Eqs.~\eqref{dc17} and \eqref{dc18}, we find the sourced massive equations
\begin{equation}\label{rsadsdc5}
    \begin{split}
    \overline{\nabla}_{\lambda}F^{\lambda \mu}(A_k) +\frac{2\Lambda}{3}A^{\mu}_k=-2\kappa_kT^{\mu}_0(g_k),
    \end{split}    
\end{equation}
\begin{equation}\label{rsadsdc6}
\overline{\nabla}^2\phi_k+\frac{2\Lambda}{3}\phi_k=-2\kappa_kT^{0}_{0}(g_k),
\end{equation}
where
\begin{equation}\label{rsadsdc7}
    T^{0}_{0}(g_k)=-\frac{Q_k^2}{8\pi \epsilon_0r^4}, \ \ \ \ \ T^{i}_{0}(g_k)=0.
\end{equation}
Equation \eqref{rsadsdc5} identifies $A^{\mu}_{k}$ as a Proca field with mass $M_k=\frac{2\Lambda}{3}$, driven by the source $T^{\mu}_{0}(g_k)$. We may regard it as a massive vector potential produced by a point charge $q_k$ and a distributed electrostatic source $\rho_k$
\begin{equation}\label{rsadsdc8}
    q_k=2C^2(k)G_km_k, \ \ \ \ \ \rho_k=\frac{Q_k^2}{4\pi r^4}.
\end{equation} 
Here $m_k$ and $G_k$ determine the point charge, $T^{0}_{0}(g_k)$ supplies the non-uniform source, and $\Lambda$ sets the Proca mass. The scalar field $\phi_k$ obeys the corresponding inhomogeneous Klein--Gordon equation with the same mass $M_k=\frac{2\Lambda}{3}$ and is sourced by the same $T^{0}_{0}(g_k)$; consequently, it acts as a massive spin-0 electrostatic potential generated by $q_k$ and $\rho_k$ as above.
\subsection{Example: Multi-Kerr}
We begin with the multigravity Multi-Kerr solution derived in Section \ref{S3}; see Eq.~\eqref{kct18} (with Kerr--Schild data in Eqs.~\eqref{eq3.1.2}--\eqref{eq3.1.3}). In this case the single- and zero-copy fields read
\begin{equation}\label{kdc4}
    A^{\mu}_j =C^2(j)\frac{2m_jr}{\Sigma} l^{\mu}, \ \ \ \ \  \phi_j=C^2(j)\frac{2m_jr}{\Sigma}.
\end{equation}
These expressions solve the equations of motion \eqref{dc17} and \eqref{dc18}, namely
\begin{equation}\label{kdc5}
    \begin{split}
    \overline{\nabla}_{\lambda}F^{\lambda \mu}(A_k) =0,
    \end{split}    
\end{equation}
\begin{equation}\label{kdc6}
\overline{\nabla}^2\phi_k=0.
\end{equation}
Hence $A^{\mu}_{k}$ satisfies the vacuum Maxwell equations and represents a massless spin-1 field, while $\phi_k$ satisfies the homogeneous wave equation and represents a massless spin-0 field. Both copies carry the Kerr parameters through the combination $C^2(k)m_k$ and the rotation parameter $a$.
\subsection{Example: Multi-Kerr-AdS}
For the AdS deformation, we use the Multi-Kerr--AdS solution in Section \ref{S3}; see Eq.~\eqref{kads16} (with background and null vector in Eqs.~\eqref{kads2}--\eqref{kads3}). The single- and zero-copy fields again take the form
\begin{equation}\label{kadsdc4}
    A^{\mu}_j =C^2(j)\frac{2m_jr}{\Sigma} l^{\mu}, \ \ \ \ \ \phi_j=C^2(j)\frac{2m_jr}{\Sigma}.
\end{equation}
They now satisfy the massive equations that follow from \eqref{dc17} and \eqref{dc18}:
\begin{equation}\label{kadsdc5}
    \begin{split}
    \overline{\nabla}_{\lambda}F^{\lambda \mu}(A_k)+\frac{2\Lambda}{3} A_{k}^{\mu} =0,
    \end{split}    
\end{equation}
\begin{equation}\label{kadsdc6}
\overline{\nabla}^2\phi_k+\frac{2\Lambda}{3}\phi_k=0.
\end{equation}
Thus $A^{\mu}_{k}$ obeys the vacuum Proca equation and $\phi_k$ obeys the homogeneous Klein--Gordon equation, both with mass $M_k=\frac{2\Lambda}{3}$. The mapping keeps the Kerr--AdS parameters: $C^2(k)m_k$ and the rotation parameter $a$ enter the copy fields, while $\Lambda$ supplies their mass.
\subsection{Example: Multi-Kerr-Newman}
We now use the Multi-Kerr--Newman multigravity solution of Section \ref{S3}; see Eq.~\eqref{knf} with the definitions in Eq.~\eqref{kn2}. The single- and zero-copy fields follow as
\begin{equation}\label{knndc4}
    A^{\mu}_j =C^2(j)\frac{1}{\Sigma}\left(2G_jm_jr-\frac{G_jQ_j^2}{4\pi\epsilon_{0j}} \right)k^{\mu},\quad \phi_j=C^2(j)\frac{1}{\Sigma}\left(2G_jm_jr-\frac{G_jQ_j^2}{4\pi\epsilon_{0j}} \right).
\end{equation}
Plugging Eq.~\eqref{knndc4} into \eqref{dc17} and \eqref{dc18} gives the equations with external source
\begin{equation}\label{knndc5}
    \begin{split}
    \overline{\nabla}_{\lambda}F^{\lambda \mu}(A_k) =-2\kappa_kT^{\mu}_0(g_k),
    \end{split}    
\end{equation}
\begin{equation}\label{knndc6}
\overline{\nabla}^2\phi_k=-2\kappa_kT^{0}_{0}(g_k),
\end{equation}
where
\begin{equation}\label{knndc7}
    T^{0}_{0}(g_k)=-\frac{Q_k^2}{8\pi \epsilon_0\Sigma^2}, \ \ \ \ \ T^{i}_{0}(g_k)=0.
\end{equation}
Therefore $A^{\mu}_{k}$ satisfies Maxwell's equations with source $T^{\mu}_{0}(g_k)$, while $\phi_k$ satisfies the corresponding inhomogeneous wave equation with source $T^{0}_{0}(g_k)$. The copy fields inherit the Kerr--Newman data through $C^2(k)m_k$, the rotation parameter $a$, the gravitational constant $G_j$, and the charge $Q_j$. In particular, the Einstein--Maxwell stress-energy component $T^{0}_{0}(g_k)$ induces the sources
\begin{equation}\label{knndc8}
J_k^{\mu}=\frac{Q_k^2}{4\pi \epsilon_0 \Sigma^2}\delta^{\mu}_{0}, \ \ \ \ \ \rho_k=\frac{Q_k^2}{4\pi \epsilon_0 \Sigma^2}.
\end{equation}
\subsection{Example: Multi-Kerr-Newman-AdS}
Finally, we consider the Multi-Kerr--Newman--AdS multigravity solution presented in Section \ref{S3}; see Eq.~\eqref{knads4} with the definitions in Eqs.~\eqref{knads2}--\eqref{knads3}. The associated copy fields are
\begin{equation}\label{knadsdc4}
    A^{\mu}_j =C^2(j)\frac{1}{\Sigma}\left(2G_jm_jr-\frac{G_jQ_j^2}{4\pi\epsilon_{0j}} \right)k^{\mu},\quad \phi_j=C^2(j)\frac{1}{\Sigma}\left(2G_jm_jr-\frac{G_jQ_j^2}{4\pi\epsilon_{0j}} \right).
\end{equation}
They obey the massive sourced equations that arise from \eqref{dc17} and \eqref{dc18}
\begin{equation}\label{knadsdc5}
    \begin{split}
    \overline{\nabla}_{\lambda}F^{\lambda \mu}(A_k) +\frac{2\Lambda}{3}A^{\mu}_k=-2\kappa_kT^{\mu}_0(g_k),
    \end{split}    
\end{equation}
\begin{equation}\label{knadsdc6}
\overline{\nabla}^2\phi_k+\frac{2\Lambda}{3}\phi_k=-2\kappa_kT^{0}_{0}(g_k),
\end{equation}
with
\begin{equation}\label{knadsdc7}
    T^{0}_{0}(g_k)=-\frac{Q_k^2}{8\pi \epsilon_0\Sigma^2},\quad T^{i}_{0}(g_k)=0.
\end{equation}
Thus $A^{\mu}_{k}$ satisfies the Proca equation with mass $M_k=\frac{2\Lambda}{3}$ in the presence of the source $T^{\mu}_{0}(g_k)$, and $\phi_k$ satisfies the corresponding inhomogeneous Klein--Gordon equation with the same mass. Both fields inherit $C^2(k)m_k$, the rotation parameter $a$, the gravitational constant $G_j$, and the charge $Q_j$ from the Kerr--Newman--AdS geometry, while $\Lambda$ fixes their mass. Moreover, the Einstein--Maxwell stress-energy component $T^{0}_{0}(g_k)$ generates the sources
\begin{equation}\label{knadsdc8}
    J_k^{\mu}=\frac{Q_k^2}{4\pi \epsilon_0 \Sigma^2}\delta^{\mu}_{0}, \ \ \ \ \ \rho_k=\frac{Q_k^2}{4\pi \epsilon_0 \Sigma^2}.
\end{equation}
\subsection{Example: Multi-Taub-NUT}
Now we proceed with the Multi-Taub--NUT multigravity solution derived in Section \ref{S4}; see Eq.~\eqref{tnads1}, obtained from the double Kerr--Schild seed \eqref{tn1}. The single and zero-th copy fields are given by
\begin{equation}\label{tndc5}
    A^{\mu}_j =C^2(j)\kappa_j\left(\frac{2N_jp}{q^2-p^2}k^{\mu}+\frac{2M_jq}{q^2-p^2}l^{\mu} \right),\quad \phi_j=C^2(j)\kappa_j\left(\frac{2N_jp}{q^2-p^2}+\frac{2M_jq}{q^2-p^2} \right).
\end{equation}
They satisfy the equations of motion \eqref{dc17} and \eqref{dc18}, which become
\begin{equation}\label{tndc6}
    \begin{split}
    \overline{\nabla}_{\lambda}F^{\lambda \mu}(A_k) +\frac{2\Lambda}{3}A^{\mu}_k=0,
    \end{split}    
\end{equation}
\begin{equation}\label{tndc7}
\overline{\nabla}^2\phi_k+\frac{2\Lambda}{3}\phi_k=0.
\end{equation}
Consequently, $A^{\mu}_{k}$ solves the vacuum Proca equation and $\phi_k$ solves the homogeneous Klein--Gordon equation, both with mass $M_k=\frac{2\Lambda}{3}$. The mapping transfers the parameters of the Taub--NUT geometry: the mass combination $C^2(k)m_k$, the NUT parameter $N_k$, and the coupling $\kappa_j$ appear in the copy fields, while $\Lambda$ sets their mass.
\subsection{Example: Multi-Plebański-Demiański}
We close this list with the Multi-Pleba\'nski--Demia\'nski multigravity solution constructed in Section \ref{S4}; see Eq.~\eqref{pds1}, built from the double Kerr--Schild seed \eqref{pd1} and the definitions in Eqs.~\eqref{pd2}--\eqref{pd4}. The corresponding copy fields are
$$
A^{\mu}_j =C^2(j)\frac{\kappa_j}{2}\left(\frac{2N_jp+G_j^2}{8\pi(q^2-p^2)}k^{\mu}+\frac{2M_jq-Q_j^2}{8\pi(q^2-p^2)}l^{\mu}\right),
$$
\begin{equation}\label{pddc5}
\phi_j=C^2(j)\frac{\kappa_j}{2}\left(\frac{2N_jp+G_j^2}{8\pi(q^2-p^2)}+\frac{2M_jq-Q_j^2}{8\pi(q^2-p^2)} \right).
\end{equation}
These fields satisfy \eqref{dc17} and \eqref{dc18}, which in this case read
\begin{equation}\label{pddc6}
    \begin{split}
    \overline{\nabla}_{\lambda}F^{\lambda \mu}(A_k) +\frac{2\Lambda}{3}A^{\mu}_k=0,
    \end{split}    
\end{equation}
\begin{equation}\label{pddc7}
\overline{\nabla}^2\phi_k+\frac{2\Lambda}{3}\phi_k=0.
\end{equation}
Thus $A^{\mu}_{k}$ obeys the vacuum Proca equation and $\phi_k$ obeys the homogeneous Klein--Gordon equation, both with mass $M_k=\frac{2\Lambda}{3}$. The copy fields carry the Plebański--Demiański parameters: the mass combination $C^2(k)m_k$, the NUT parameter $N_k$, the magnetic and electric charges $G_k$ and $Q_k$, and the coupling $\kappa_j$ enter explicitly, while $\Lambda$ again supplies the field mass.

\section{Final Remarks}\label{S6}
In this work we have identified additional new solutions in the context of multimetric gravity. By varying the action for multigravity \eqref{MG1} and setting this variation to zero, we derived the general equations of motion for the $N$ interacting spin-2 fields $(g_j)_{\mu \nu}$, as presented in equation \eqref{eom10}. From this dynamics, we initiated an investigation to construct exact solutions by focusing on the study of proportional Kerr-Schild ansatz. We derived the dynamics for the proportional Kerr-Schild ansatz as shown in equation \eqref{eq3.1.12}, which led to the re-derivation in the metric formalism of the two axial-symmetric solutions in multigravity: Multi-Kerr and Multi-Kerr-AdS. These solutions were verified to coincide with those found using the vielbein formalism in \cite{Wood:2024acv} and in particular to coincide in the $N=2$ bigravity scenario solutions found in \cite{Ayon-Beato:2015qtt}.

Following our initial exploration, we focused on examining the dynamics of multigravity specifically for proportional Kerr-Schild spacetimes. This investigation allowed us to develop a general approach for constructing constant Ricci scalar proportional Kerr-Schild solutions within multigravity, utilizing spin-2 metric fields that satisfy Einstein's equations along with additional constraints outlined in equation \eqref{ksextra}. Through this approach, we were able to derive all the exact solutions presented in Section \ref{S3}. We re-derive in the metric formalism of multigravity the black hole solutions previously found in \cite{Wood:2024acv} including the Multi-Kerr-Newman-AdS black hole in multigravity and all its limiting cases as well such as Schwarzschild, Schwarzschild-AdS, Kerr, Kerr-AdS, Reissner-Nordström and Reissner-Nordström-AdS. Furthermore, we obtained new solutions in Kerr-Schild form in multigravity that encompass Kundt waves, $pp$-waves, and Siklos-AdS waves. In this part, we concluded our study of exact solutions in multigravity by investigating the Proportional Double Kerr-Schild ansatz, leading to the derivation of the Multi-Taub-NUT and Multi-Plebański-Demiański solutions in multigravity.

In the end, after constructing Kerr-Schild and Double-Kerr-Schild type solutions within the framework of multigravity, we explored the Classical Double Copy relationships in this context for stationary Kerr-Schild and double-Kerr-Schild spacetimes. We obtained the dynamics for the single copy fields $A^{\mu}_k$, which are derived from the spin-2 metric fields $(g_k)_{\mu \nu}$ and still obey Maxwell or Proca equations, as well as for the zero copy fields $\phi_k$, which obey the wave or Klein-Gordon equations, demonstrating that within this multigravity framework, the classical double copy method remains applicable to find gauge theory solutions from multigravity solutions.

In future research, we identify two interconnected directions: the development of more exact solutions in multigravity is essential, extending beyond just the Proportional Kerr-Schild or double Kerr-Schild models to broaden these findings. Some work on non-proportional solutions in the context of multigravity has been explored in \cite{Wood:2024eol,Brizuela:2025ldm}. From a double copy standpoint, investigating the Weyl double copy for Type {\bf D} and Type {\bf N} spacetimes within the multigravity framework would be interesting and a study of how the instabilities of the multigravity solutions will reflect in the single and zero-copies descriptions would be very important.

\acknowledgments 
It is a pleasure to thank César Ramos for useful discussions and commments.
Everardo Rivera-Oliva wishes to express his gratitude for the support provided by the Ph.D. scholarship No. 743129 from SECIHTI (CONAHCYT)-Mexico.

\vskip 2truecm


\begin{thebibliography}{99}

\bibitem{Armas:2021yut}
J.~Armas,
``Conversations on Quantum Gravity,''
Cambridge University Press, 2021,
ISBN 978-1-316-71763-9, 978-1-107-16887-9
doi:10.1017/9781316717639

\bibitem{Feynman:1996kb}
R.~P.~Feynman, F.~B.~Morinigo, W.~G.~Wagner and B.~Hatfield,
``Feynman lectures on gravitation,''
doi:10.1201/9780429502859

\bibitem{Gupta:1954zz}
S.~N.~Gupta,
``Gravitation and Electromagnetism,''
Phys. Rev. \textbf{96}, 1683-1685 (1954)
doi:10.1103/PhysRev.96.1683

\bibitem{Wald:1986bj}
R.~M.~Wald,
``Spin-2 Fields and General Covariance,''
Phys. Rev. D \textbf{33}, 3613 (1986)
doi:10.1103/PhysRevD.33.3613

\bibitem{Weinberg:1965rz}
S.~Weinberg,
``Photons and gravitons in perturbation theory: Derivation of Maxwell's and Einstein's equations,''
Phys. Rev. \textbf{138}, B988-B1002 (1965)
doi:10.1103/PhysRev.138.B988

\bibitem{Weinberg:1980kq}
S.~Weinberg and E.~Witten,
``Limits on Massless Particles,''
Phys. Lett. B \textbf{96}, 59-62 (1980)
doi:10.1016/0370-2693(80)90212-9

\bibitem{Mattingly:2005re}
D.~Mattingly,
``Modern tests of Lorentz invariance,''
Living Rev. Rel. \textbf{8}, 5 (2005)
doi:10.12942/lrr-2005-5
[arXiv:gr-qc/0502097 [gr-qc]].

\bibitem{Deser:2007jk}
S.~Deser and R.~P.~Woodard,
``Nonlocal Cosmology,''
Phys. Rev. Lett. \textbf{99}, 111301 (2007)
doi:10.1103/PhysRevLett.99.111301
[arXiv:0706.2151 [astro-ph]].

\bibitem{Fierz:1939ix}
M.~Fierz and W.~Pauli,
``On relativistic wave equations for particles of arbitrary spin in an electromagnetic field,''
Proc. Roy. Soc. Lond. A \textbf{173}, 211-232 (1939)
doi:10.1098/rspa.1939.0140

\bibitem{vanDam:1970vg}
H.~van Dam and M.~J.~G.~Veltman,
``Massive and massless Yang-Mills and gravitational fields,''
Nucl. Phys. B \textbf{22}, 397-411 (1970)
doi:10.1016/0550-3213(70)90416-5

\bibitem{Zakharov:1970cc}
V.~I.~Zakharov,
``Linearized gravitation theory and the graviton mass,''
JETP Lett. \textbf{12}, 312 (1970)

\bibitem{Boulware:1972yco}
D.~G.~Boulware and S.~Deser,
``Can gravitation have a finite range?,''
Phys. Rev. D \textbf{6}, 3368-3382 (1972)
doi:10.1103/PhysRevD.6.3368

\bibitem{VanNieuwenhuizen:1973fi}
P.~Van Nieuwenhuizen,
``On ghost-free tensor lagrangians and linearized gravitation,''
Nucl. Phys. B \textbf{60}, 478-492 (1973)
doi:10.1016/0550-3213(73)90194-6

\bibitem{deRham:2010kj}
C.~de Rham, G.~Gabadadze and A.~J.~Tolley,
``Resummation of Massive Gravity,''
Phys. Rev. Lett. \textbf{106}, 231101 (2011)
doi:10.1103/PhysRevLett.106.231101
[arXiv:1011.1232 [hep-th]].

\bibitem{Vainshtein:1972sx}
A.~I.~Vainshtein,
``To the problem of nonvanishing gravitation mass,''
Phys. Lett. B \textbf{39}, 393-394 (1972)
doi:10.1016/0370-2693(72)90147-5

\bibitem{Hassan:2011zd}
S.~F.~Hassan and R.~A.~Rosen,
``Bimetric Gravity from Ghost-free Massive Gravity,''
JHEP \textbf{02}, 126 (2012)
doi:10.1007/JHEP02(2012)126
[arXiv:1109.3515 [hep-th]].


\bibitem{Hinterbichler:2011tt}
K.~Hinterbichler,
``Theoretical Aspects of Massive Gravity,''
Rev. Mod. Phys. \textbf{84}, 671-710 (2012)
doi:10.1103/RevModPhys.84.671
[arXiv:1105.3735 [hep-th]].

\bibitem{deRham:2014zqa}
C.~de Rham,
``Massive Gravity,''
Living Rev. Rel. \textbf{17}, 7 (2014)
doi:10.12942/lrr-2014-7
[arXiv:1401.4173 [hep-th]].

\bibitem{Schmidt-May:2015vnx}
A.~Schmidt-May and M.~von Strauss,
``Recent developments in bimetric theory,''
J. Phys. A \textbf{49}, no.18, 183001 (2016)
doi:10.1088/1751-8113/49/18/183001
[arXiv:1512.00021 [hep-th]].

\bibitem{intro1}
S. Perlmutter, G. Aldering, G. Goldhaber, R. A. Knop, P. Nugent, P. G. Castro, S. Deustua,
S. Fabbro, A. Goobar, D. E. Groom, I. M. Hook, A. G. Kim, M. Y. Kim, J. C. Lee, N. J.
Nunes, R. Pain, C. R. Pennypacker, R. Quimby, C. Lidman, R. S. Ellis, M. Irwin, R. G.
McMahon, P. Ruiz-Lapuente, N. Walton, B. Schaefer, B. J. Boyle, A. V. Filippenko,
T. Matheson, A. S. Fruchter, N. Panagia, H. J. M. Newberg, W. J. Couch, and The
Supernova Cosmology Project. Measurements of $\omega$ and $\lambda$ from 42 high-redshift supernovae. The Astrophysical Journal, 517(2):565–586, June 1999.

\bibitem{intro2}
Adam G. Riess, Alexei V. Filippenko, Peter Challis, Alejandro Clocchiatti, Alan Diercks,
Peter M. Garnavich, Ron L. Gilliland, Craig J. Hogan, Saurabh Jha, Robert P. Kirshner,
B. Leibundgut, M. M. Phillips, David Reiss, Brian P. Schmidt, Robert A. Schommer,
R. Chris Smith, J. Spyromilio, Christopher Stubbs, Nicholas B. Suntzeff, and John Tonry.
Observational evidence from supernovae for an accelerating universe and a cosmological
constant. The Astronomical Journal, 116(3):1009–1038, September 1998.

\bibitem{GonzalezAlbornoz:2017gbh}
N.~L.~Gonz\'alez Albornoz, A.~Schmidt-May and M.~von Strauss,
``Dark matter scenarios with multiple spin-2 fields,''
JCAP \textbf{01}, 014 (2018)
doi:10.1088/1475-7516/2018/01/014
[arXiv:1709.05128 [hep-th]].

\bibitem{Babichev:2016bxi}
E.~Babichev, L.~Marzola, M.~Raidal, A.~Schmidt-May, F.~Urban, H.~Veerm\"ae and M.~von Strauss,
``Heavy spin-2 Dark Matter,''
JCAP \textbf{09}, 016 (2016)
doi:10.1088/1475-7516/2016/09/016
[arXiv:1607.03497 [hep-th]].

\bibitem{Babichev:2016hir}
E.~Babichev, L.~Marzola, M.~Raidal, A.~Schmidt-May, F.~Urban, H.~Veerm{\"a}e and M.~von Strauss,
``Bigravitational origin of dark matter,''
Phys. Rev. D \textbf{94}, no.8, 084055 (2016)
doi:10.1103/PhysRevD.94.084055
[arXiv:1604.08564 [hep-ph]].

\bibitem{Marzola:2017lbt}
L.~Marzola, M.~Raidal and F.~R.~Urban,
``Oscillating Spin-2 Dark Matter,''
Phys. Rev. D \textbf{97}, no.2, 024010 (2018)
doi:10.1103/PhysRevD.97.024010
[arXiv:1708.04253 [hep-ph]].

\bibitem{Niedermann:2018lhx}
F.~Niedermann, A.~Padilla and P.~M.~Saffin,
``Higher Order Clockwork Gravity,''
Phys. Rev. D \textbf{98}, no.10, 104014 (2018)
doi:10.1103/PhysRevD.98.104014
[arXiv:1805.03523 [hep-th]].

\bibitem{Avgoustidis:2020wrd}
A.~Avgoustidis, F.~Niedermann, A.~Padilla and P.~M.~Saffin,
``Deconstructing higher order clockwork gravity,''
Phys. Rev. D \textbf{103}, no.12, 124007 (2021)
doi:10.1103/PhysRevD.103.124007
[arXiv:2010.10970 [hep-th]].

\bibitem{Hassan:2012wr}
S.~F.~Hassan, A.~Schmidt-May and M.~von Strauss,
``On Consistent Theories of Massive Spin-2 Fields Coupled to Gravity,''
JHEP \textbf{05}, 086 (2013)
doi:10.1007/JHEP05(2013)086
[arXiv:1208.1515 [hep-th]].

\bibitem{Baldacchino:2016jsz}
O.~Baldacchino and A.~Schmidt-May,
``Structures in multiple spin-2 interactions,''
J. Phys. A \textbf{50}, no.17, 175401 (2017)
doi:10.1088/1751-8121/aa649d
[arXiv:1604.04354 [gr-qc]].

\bibitem{Babichev:2013una}
E.~Babichev and A.~Fabbri,
``Instability of black holes in massive gravity,''
Class. Quant. Grav. \textbf{30}, 152001 (2013)
doi:10.1088/0264-9381/30/15/152001
[arXiv:1304.5992 [gr-qc]].

\bibitem{Brito:2013wya}
R.~Brito, V.~Cardoso and P.~Pani,
Phys. Rev. D \textbf{88}, no.2, 023514 (2013)
doi:10.1103/PhysRevD.88.023514
[arXiv:1304.6725 [gr-qc]].

\bibitem{Babichev:2014oua}
E.~Babichev and A.~Fabbri,
``Stability analysis of black holes in massive gravity: a unified treatment,''
Phys. Rev. D \textbf{89}, no.8, 081502 (2014)
doi:10.1103/PhysRevD.89.081502
[arXiv:1401.6871 [gr-qc]].

\bibitem{Babichev:2015zub}
E.~Babichev, R.~Brito and P.~Pani,
``Linear stability of nonbidiagonal black holes in massive gravity,''
Phys. Rev. D \textbf{93}, no.4, 044041 (2016)
doi:10.1103/PhysRevD.93.044041
[arXiv:1512.04058 [gr-qc]].

\bibitem{Babichev:2015xha}
E.~Babichev and R.~Brito,
``Black holes in massive gravity,''
Class. Quant. Grav. \textbf{32}, 154001 (2015)
doi:10.1088/0264-9381/32/15/154001
[arXiv:1503.07529 [gr-qc]].
\bibitem{Bern:2008qj}
Z.~Bern, J.~J.~M.~Carrasco and H.~Johansson,
``New Relations for Gauge-Theory Amplitudes,''
Phys. Rev. D \textbf{78}, 085011 (2008)
doi:10.1103/PhysRevD.78.085011
[arXiv:0805.3993 [hep-ph]].

\bibitem{Bern:2010ue}
Z.~Bern, J.~J.~M.~Carrasco and H.~Johansson,
``Perturbative Quantum Gravity as a Double Copy of Gauge Theory,''
Phys. Rev. Lett. \textbf{105}, 061602 (2010)
doi:10.1103/PhysRevLett.105.061602
[arXiv:1004.0476 [hep-th]].

\bibitem{Bern:2010yg}
Z.~Bern, T.~Dennen, Y.~t.~Huang and M.~Kiermaier,
``Gravity as the Square of Gauge Theory,''
Phys. Rev. D \textbf{82}, 065003 (2010)
doi:10.1103/PhysRevD.82.065003
[arXiv:1004.0693 [hep-th]].

\bibitem{Kawai:1985xq}
H.~Kawai, D.~C.~Lewellen and S.~H.~H.~Tye,
``A Relation Between Tree Amplitudes of Closed and Open Strings,''
Nucl. Phys. B \textbf{269} (1986), 1-23
doi:10.1016/0550-3213(86)90362-7

\bibitem{Bianchi:2008pu}
M.~Bianchi, H.~Elvang and D.~Z.~Freedman,
``Generating Tree Amplitudes in N=4 SYM and N = 8 SG,''
JHEP \textbf{09}, 063 (2008)
doi:10.1088/1126-6708/2008/09/063
[arXiv:0805.0757 [hep-th]].

\bibitem{Bern:1998sv}
Z.~Bern, L.~J.~Dixon, M.~Perelstein and J.~S.~Rozowsky,
``Multileg one loop gravity amplitudes from gauge theory,''
Nucl. Phys. B \textbf{546}, 423-479 (1999)
doi:10.1016/S0550-3213(99)00029-2
[arXiv:hep-th/9811140 [hep-th]].

\bibitem{Borsten:2020zgj}
L.~Borsten, B.~Jur{\v{c}}o, H.~Kim, T.~Macrelli, C.~Saemann and M.~Wolf,
``Becchi-Rouet-Stora-Tyutin-Lagrangian Double Copy of Yang-Mills Theory,''
Phys. Rev. Lett. \textbf{126}, no.19, 191601 (2021)
doi:10.1103/PhysRevLett.126.191601
[arXiv:2007.13803 [hep-th]].

\bibitem{Borsten:2021gyl}
L.~Borsten, H.~Kim, B.~Jur{\v{c}}o, T.~Macrelli, C.~Saemann and M.~Wolf,
``Tree-level color{\textendash}kinematics duality implies loop-level color{\textendash}kinematics duality up to counterterms,''
Nucl. Phys. B \textbf{989}, 116144 (2023)
doi:10.1016/j.nuclphysb.2023.116144
[arXiv:2108.03030 [hep-th]].

\bibitem{White:2017mwc}
C.~D.~White,
``The double copy: gravity from gluons,''
Contemp. Phys. \textbf{59}, 109 (2018)
doi:10.1080/00107514.2017.1415725
[arXiv:1708.07056 [hep-th]].
\bibitem{Monteiro:2014cda}
R.~Monteiro, D.~O'Connell and C.~D.~White,
``Black holes and the double copy,''
JHEP \textbf{12}, 056 (2014)
doi:10.1007/JHEP12(2014)056
[arXiv:1410.0239 [hep-th]].

\bibitem{White:2024pve}
C.~D.~White,
``The Classical Double Copy,''
World Scientific, 2024,
ISBN 978-1-80061-545-8, 978-1-80061-547-2
doi:10.1142/q0457

\bibitem{Kerr:1965wfc}
R.~P.~Kerr and A.~Schild,
``Some algebraically degenerate solutions of Einstein\textquoteright{}s gravitational field equations,''
Proc. Symp. Appl. Math. \textbf{17}, 199 (1965)

\bibitem{Luna:2015paa}
A.~Luna, R.~Monteiro, D.~O'Connell and C.~D.~White,
``The classical double copy for Taub{\textendash}NUT spacetime,''
Phys. Lett. B \textbf{750}, 272-277 (2015)
doi:10.1016/j.physletb.2015.09.021
[arXiv:1507.01869 [hep-th]].

\bibitem{Luna:2016due}
A.~Luna, R.~Monteiro, I.~Nicholson, D.~O'Connell and C.~D.~White,
``The double copy: Bremsstrahlung and accelerating black holes,''
JHEP \textbf{06}, 023 (2016)
doi:10.1007/JHEP06(2016)023
[arXiv:1603.05737 [hep-th]].

\bibitem{Ridgway:2015fdl}
A.~K.~Ridgway and M.~B.~Wise,
``Static Spherically Symmetric Kerr-Schild Metrics and Implications for the Classical Double Copy,''
Phys. Rev. D \textbf{94}, no.4, 044023 (2016)
doi:10.1103/PhysRevD.94.044023
[arXiv:1512.02243 [hep-th]].

\bibitem{Bahjat-Abbas:2017htu}
N.~Bahjat-Abbas, A.~Luna and C.~D.~White,
``The Kerr-Schild double copy in curved spacetime,''
JHEP \textbf{12}, 004 (2017)
doi:10.1007/JHEP12(2017)004
[arXiv:1710.01953 [hep-th]].

\bibitem{CarrilloGonzalez:2019gof}
M.~Carrillo Gonz{\'a}lez, B.~Melcher, K.~Ratliff, S.~Watson and C.~D.~White,
``The classical double copy in three spacetime dimensions,''
JHEP \textbf{07}, 167 (2019)
doi:10.1007/JHEP07(2019)167
[arXiv:1904.11001 [hep-th]].

\bibitem{Berman:2018hwd}
D.~S.~Berman, E.~Chac{\'o}n, A.~Luna and C.~D.~White,
``The self-dual classical double copy, and the Eguchi-Hanson instanton,''
JHEP \textbf{01}, 107 (2019)
doi:10.1007/JHEP01(2019)107
[arXiv:1809.04063 [hep-th]].

\bibitem{Carrillo-Gonzalez:2017iyj}
M.~Carrillo-Gonz{\'a}lez, R.~Penco and M.~Trodden,
``The classical double copy in maximally symmetric spacetimes,''
JHEP \textbf{04}, 028 (2018)
doi:10.1007/JHEP04(2018)028
[arXiv:1711.01296 [hep-th]].

\bibitem{Luna:2018dpt}
A.~Luna, R.~Monteiro, I.~Nicholson and D.~O'Connell,
``Type D Spacetimes and the Weyl Double Copy,''
Class. Quant. Grav. \textbf{36}, 065003 (2019)
doi:10.1088/1361-6382/ab03e6
[arXiv:1810.08183 [hep-th]].

\bibitem{White:2020sfn}
C.~D.~White,
``Twistorial Foundation for the Classical Double Copy,''
Phys. Rev. Lett. \textbf{126}, no.6, 061602 (2021)
doi:10.1103/PhysRevLett.126.061602
[arXiv:2012.02479 [hep-th]].

\bibitem{Chacon:2021wbr}
E.~Chac{\'o}n, S.~Nagy and C.~D.~White,
``The Weyl double copy from twistor space,''
JHEP \textbf{05}, 2239 (2021)
doi:10.1007/JHEP05(2021)239
[arXiv:2103.16441 [hep-th]].

\bibitem{Chacon:2021lox}
E.~Chac{\'o}n, S.~Nagy and C.~D.~White,
``Alternative formulations of the twistor double copy,''
JHEP \textbf{03}, 180 (2022)
doi:10.1007/JHEP03(2022)180
[arXiv:2112.06764 [hep-th]].

\bibitem{Luna:2022dxo}
A.~Luna, N.~Moynihan and C.~D.~White,
``Why is the Weyl double copy local in position space?,''
JHEP \textbf{12}, 046 (2022)
doi:10.1007/JHEP12(2022)046
[arXiv:2208.08548 [hep-th]].

\bibitem{Garcia-Compean:2024uie}
H.~Garc{\'\i}a-Compe{\'a}n and C.~Ramos,
``Classical Kerr-Schild double copy in bigravity for maximally symmetric spacetimes,''
JHEP \textbf{07}, 074 (2024)
doi:10.1007/JHEP07(2024)074
[arXiv:2403.19608 [gr-qc]].

\bibitem{Garcia-Compean:2024zze}
H.~Garc{\'\i}a-Compe{\'a}n and C.~I.~Ramos,
``Pleba{\'n}ski-Demia{\'n}ski solutions in bigravity and Kerr-Schild double copy relations using an effective metric,''
JHEP \textbf{06}, 079 (2025)
doi:10.1007/JHEP06(2025)079
[arXiv:2412.17191 [gr-qc]].

\bibitem{Garcia-Compean:2025wkj}
H.~Garc{\'\i}a-Compe{\'a}n and C.~I.~Ramos,
``Weyl double copy in bimetric massive gravity,''
JHEP \textbf{11}, 154 (2025)
doi:10.1007/JHEP11(2025)154
[arXiv:2510.01550 [gr-qc]].
\bibitem{Fabbrichesi2021}
M.~Fabbrichesi, E.~Gabrielli, and G.~Lanfranchi, 
\textit{The Physics of the Dark Photon: A Primer}, 
SpringerBriefs in Physics, Springer International Publishing, 2021. 
doi:10.1007/978-3-030-62519-1

\bibitem{PhysRevD.110.L041302}
L.~Pinol,
\textit{Effective field theory of multifield inflationary fluctuations},
Phys. Rev. D \textbf{110}, L041302 (2024).
doi:10.1103/PhysRevD.110.L041302

\bibitem{Wands2007}
D.~Wands, 
\textit{Multiple Field Inflation}, 
en: M.~Lemoine, J.~Martin, P.~Peter (eds), 
\textit{Inflationary Cosmology}, 
Lecture Notes in Physics, vol 738, 
Springer, Berlin, Heidelberg, 2007, pp. 275--304.
doi:10.1007/978-3-540-74353-8\_8


\bibitem{Wood:2024acv}
K.~Wood, P.~M.~Saffin and A.~Avgoustidis,
``Black holes in multimetric gravity,''
Phys. Rev. D \textbf{109}, no.12, 124006 (2024)
doi:10.1103/PhysRevD.109.124006
[arXiv:2402.17835 [gr-qc]].

\bibitem{Chamseddine:2011mu}
A.~H.~Chamseddine and V.~Mukhanov,
``Massive Gravity Simplified: A Quadratic Action,''
JHEP \textbf{08}, 091 (2011)
doi:10.1007/JHEP08(2011)091
[arXiv:1106.5868 [hep-th]].

\bibitem{GrootNibbelink:2006vzk}
S.~Groot Nibbelink, M.~Peloso and M.~Sexton,
``Nonlinear Properties of Vielbein Massive Gravity,''
Eur. Phys. J. C \textbf{51}, 741-752 (2007)
doi:10.1140/epjc/s10052-007-0311-x
[arXiv:hep-th/0610169 [hep-th]].

\bibitem{Hinterbichler:2012cn}
K.~Hinterbichler and R.~A.~Rosen,
``Interacting Spin-2 Fields,''
JHEP \textbf{07}, 047 (2012)
doi:10.1007/JHEP07(2012)047
[arXiv:1203.5783 [hep-th]].

\bibitem{Zumino:1970tu}
B.~Zumino,
``Effective Lagrangians and Broken Symmetries,'' In Lectures on Elementary
Particles and Quantum Field Theory, volume 2, page 437, January 1970.

\bibitem{Israelit:1986ez}
M.~Israelit and N.~Rosen,
``A Gauge Covariant Bimetric Theory of Gravitation and Electromagnetism,''
Found. Phys. \textbf{19}, 33-55 (1989)
doi:10.1007/BF00737765

\bibitem{Rosen:1978mb}
N.~Rosen,
``Bimetric Gravitation Theory and PSR-1913+16,''
Astrophys. J. \textbf{221}, 284-285 (1978)
doi:10.1086/156027

\bibitem{Kluson:2013jlo}
J.~Kluson,
``Is Bimetric Gravity Really Ghost Free?,''
Int. J. Mod. Phys. A \textbf{28}, no.28, 1350143 (2013)
doi:10.1142/S0217751X13501431
[arXiv:1301.3296 [hep-th]].

\bibitem{Kluson:2012ps}
J.~Kluso{\v{n}},
``Hamiltonian Formalism of Particular Bimetric Gravity Model,''
Phys. Rev. D \textbf{87}, no.8, 084017 (2013)
doi:10.1103/PhysRevD.87.084017
[arXiv:1211.6267 [hep-th]].

\bibitem{Kluson:2013lza}
J.~Kluson,
``Hamiltonian Formalism of General Bimetric Gravity,''
Eur. Phys. J. C \textbf{73}, no.9, 2553 (2013)
doi:10.1140/epjc/s10052-013-2553-0
[arXiv:1303.1652 [hep-th]].

\bibitem{Kluson:2013aca}
J.~Kluso{\v{n}},
``Hamiltonian Formalism of Bimetric Gravity In Vierbein Formulation,''
Eur. Phys. J. C \textbf{74}, no.8, 2985 (2014)
doi:10.1140/epjc/s10052-014-2985-1
[arXiv:1307.1974 [hep-th]].

\bibitem{Damour:2002wu}
T.~Damour, I.~I.~Kogan and A.~Papazoglou,
``Nonlinear bigravity and cosmic acceleration,''
Phys. Rev. D \textbf{66}, 104025 (2002)
doi:10.1103/PhysRevD.66.104025
[arXiv:hep-th/0206044 [hep-th]].

\bibitem{Hassan:2011ea}
S.~F.~Hassan and R.~A.~Rosen,
``Confirmation of the Secondary Constraint and Absence of Ghost in Massive Gravity and Bimetric Gravity,''
JHEP \textbf{04}, 123 (2012)
doi:10.1007/JHEP04(2012)123
[arXiv:1111.2070 [hep-th]].

\bibitem{Hassan:2011hr}
S.~F.~Hassan and R.~A.~Rosen,
``Resolving the Ghost Problem in non-Linear Massive Gravity,''
Phys. Rev. Lett. \textbf{108}, 041101 (2012)
doi:10.1103/PhysRevLett.108.041101
[arXiv:1106.3344 [hep-th]].

\bibitem{Hassan:2012qv}
S.~F.~Hassan, A.~Schmidt-May and M.~von Strauss,
``Proof of Consistency of Nonlinear Massive Gravity in the St{\"u}ckelberg Formulation,''
Phys. Lett. B \textbf{715}, 335-339 (2012)
doi:10.1016/j.physletb.2012.07.018
[arXiv:1203.5283 [hep-th]].

\bibitem{Kluson:2012wf}
J.~Kluson,
``Non-Linear Massive Gravity with Additional Primary Constraint and Absence of Ghosts,''
Phys. Rev. D \textbf{86}, 044024 (2012)
doi:10.1103/PhysRevD.86.044024
[arXiv:1204.2957 [hep-th]].

\bibitem{Kluson:2012gz}
J.~Kluson,
``Remark About Hamiltonian Formulation of Non-Linear Massive Gravity in Stuckelberg Formalism,''
Phys. Rev. D \textbf{86}, 124005 (2012)
doi:10.1103/PhysRevD.86.124005
[arXiv:1202.5899 [hep-th]].

\bibitem{Ayon-Beato:2015qtt}
E.~Ay\'on-Beato, D.~Higuita-Borja and J.~A.~M\'endez-Zavaleta,
``Rotating (A)dS black holes in bigravity,''
Phys. Rev. D \textbf{93}, no.2, 024049 (2016)
doi:10.1103/PhysRevD.93.024049
[arXiv:1511.01108 [hep-th]].

\bibitem{Stephani:2003tm}
H.~Stephani, D.~Kramer, M.~A.~H.~MacCallum, C.~Hoenselaers and E.~Herlt,
``Exact solutions of Einstein's field equations,''
Cambridge Univ. Press, 2003,
ISBN 978-0-521-46702-5, 978-0-511-05917-9
doi:10.1017/CBO9780511535185

\bibitem{siklos1985lobatchevski}
Siklos, S. T. C. "Lobatchevski plane gravitational waves." Essays presented to WB Bonnor on his 65th birthday (1985): 247-274.

\bibitem{Ayon-Beato:2018hxz}
E.~Ay{\'o}n-Beato, D.~Higuita-Borja, J.~A.~M{\'e}ndez-Zavaleta and G.~Vel{\'a}zquez-Rodr{\'\i}guez,
``Exact ghost-free bigravitational waves,''
Phys. Rev. D \textbf{97}, no.8, 084045 (2018)
doi:10.1103/PhysRevD.97.084045
[arXiv:1801.06764 [hep-th]].

\bibitem{Bah:2019sda}
I.~Bah, R.~Dempsey and P.~Weck,
``Kerr-Schild Double Copy and Complex Worldlines,''
JHEP \textbf{02}, 180 (2020)
doi:10.1007/JHEP02(2020)180
[arXiv:1910.04197 [hep-th]].

\bibitem{Ett:2015fhw}
B.~Ett,
``Exact Solutions in Gravity: A journey through spacetime with the Kerr-Schild ansatz,''
doi:10.7275/7504405.0

\cite{Hawking:1973uf}
\bibitem{Hawking:1973uf}
S.~W.~Hawking and G.~F.~R.~Ellis,
``The Large Scale Structure of Space-Time,''
Cambridge University Press, 2023,
ISBN 978-1-009-25316-1, 978-1-009-25315-4, 978-0-521-20016-5, 978-0-521-09906-6, 978-0-511-82630-6, 978-0-521-09906-6
doi:10.1017/9781009253161

\bibitem{Newman:1965my}
E.~T.~Newman, E.~Couch, K.~Chinnapared, A.~Exton, A.~Prakash and R.~Torrence,
``Metric of a Rotating, Charged Mass,''
J. Math. Phys. \textbf{6}, 918-919 (1965)
doi:10.1063/1.1704351

\bibitem{Debney:1969zz}
G.~C.~Debney, R.~P.~Kerr and A.~Schild,
``Solutions of the Einstein and Einstein-Maxwell Equations,''
J. Math. Phys. \textbf{10}, 1842 (1969)
doi:10.1063/1.1664769

\bibitem{Gibbons:2004uw}
G.~W.~Gibbons, H.~Lu, D.~N.~Page and C.~N.~Pope,
``The General Kerr-de Sitter metrics in all dimensions,''
J. Geom. Phys. \textbf{53}, 49-73 (2005)
doi:10.1016/j.geomphys.2004.05.001
[arXiv:hep-th/0404008 [hep-th]].

\bibitem{Malek:2010mh}
T.~Malek and V.~Pravda,
``Kerr-Schild spacetimes with (A)dS background,''
Class. Quant. Grav. \textbf{28}, 125011 (2011)
doi:10.1088/0264-9381/28/12/125011
[arXiv:1009.1727 [gr-qc]].

\bibitem{Taub:1950ez}
A.~H.~Taub,
``Empty space-times admitting a three parameter group of motions,''
Annals Math. \textbf{53}, 472-490 (1951)
doi:10.2307/1969567

\bibitem{Newman:1963yy}
E.~Newman, L.~Tamburino and T.~Unti,
``Empty space generalization of the Schwarzschild metric,''
J. Math. Phys. \textbf{4}, 915 (1963)
doi:10.1063/1.1704018

\bibitem{Plebanski:1975xfb}
J.~F.~Pleba{\'n}ski,
``A class of solutions of Einstein-Maxwell equations,''
Annals Phys. \textbf{90}, no.1, 196-255 (1975)
doi:10.1016/0003-4916(75)90145-1

\bibitem{Plebanski:1976gy}
J.~F.~Pleba\'nski and M.~Demia\'nski,
``Rotating, charged, and uniformly accelerating mass in general relativity,''
Annals Phys. \textbf{98}, 98-127 (1976)
doi:10.1016/0003-4916(76)90240-2

\bibitem{Wood:2024eol}
K.~Wood, P.~M.~Saffin and A.~Avgoustidis,
``Black holes in multimetric gravity. II. Hairy solutions and linear stability of the non- and partially proportional branches,''
Phys. Rev. D \textbf{111}, no.2, 024057 (2025)
doi:10.1103/PhysRevD.111.024057
[arXiv:2410.10976 [gr-qc]].

\bibitem{Brizuela:2025ldm}
D.~Brizuela, M.~de Cesare and A.~Soler Oficial,
``Dipolar perturbations of nonbidiagonal black holes in bigravity,''
Phys. Rev. D \textbf{111}, no.8, 084084 (2025)
doi:10.1103/PhysRevD.111.084084
[arXiv:2501.16984 [gr-qc]].
\end{thebibliography}
\end{document}